\documentclass[%
 reprint,aps,pra,superscriptaddress
]{revtex4-2}

\usepackage{amsmath}
\usepackage{amssymb}
\usepackage{amsfonts}
\usepackage{mathrsfs}
\usepackage{amsthm}
\usepackage{mathtools}
\usepackage{braket}
\usepackage{physics}
\usepackage{nicematrix}
\usepackage{bm} 

\theoremstyle{definition}

\newtheorem{defin}{Definition}

\usepackage{graphicx}
\usepackage{dcolumn}
\usepackage{bm}
\usepackage{appendix}
\usepackage{amsfonts}
\usepackage[normalem]{ulem}
\usepackage{xcolor}
\usepackage{booktabs}
\usepackage{comment}
\usepackage{afterpage}

\begin{document}

\title{Characterizing randomness in parameterized quantum circuits through expressibility and average entanglement}
\author{Guilherme Ilário Correr}
\email{guilhermecorrer27@gmail.com}
\affiliation{
 Instituto de Física de São Carlos, Universidade de São Paulo, CP 369, 13560-970, São Carlos, SP, Brasil.
 }
 \affiliation{
 QTF Centre of Excellence and Department of Physics, University of Helsinki, FIN-00014 Helsinki, Finland
 }
\author{Ivan Medina}
\affiliation{
 Instituto de Física de São Carlos, Universidade de São Paulo, CP 369, 13560-970, São Carlos, SP, Brasil.
 }
\affiliation{School of Physics, Trinity College Dublin, Dublin 2, Ireland}
\author{Pedro C. Azado}
\affiliation{
 Instituto de Física de São Carlos, Universidade de São Paulo, CP 369, 13560-970, São Carlos, SP, Brasil.
 }
\author{Alexandre Drinko}
\affiliation{
 Instituto de Física de São Carlos, Universidade de São Paulo, CP 369, 13560-970, São Carlos, SP, Brasil.
 }
\author{Diogo O. Soares-Pinto}
\email{dosp@ifsc.usp.br}
\affiliation{
 Instituto de Física de São Carlos, Universidade de São Paulo, CP 369, 13560-970, São Carlos, SP, Brasil.
 }

\begin{abstract}
While scalable error correction schemes and fault tolerant quantum computing seem not to be universally accessible in the near sight, the efforts of many researchers have been directed to the exploration of the contemporary available quantum hardware. Due to these limitations, the depth and dimension of the possible quantum circuits are restricted. This motivates the study of circuits with parameterized operations that can be classically optimized in hybrid methods as variational quantum algorithms (VQAs), enabling the reduction of circuit depth and size. The characteristics of these Parameterized Quantum Circuits (PQCs) are still not fully understood outside the scope of their principal application, motivating the study of their intrinsic properties. In this work, we analyse the generation of random states in PQCs under restrictions on the qubits connectivities, justified by different quantum computer architectures. We apply the expressibility quantifier and the average entanglement as diagnostics for the characteristics of the generated states and classify the circuits depending on the topology of the quantum computer where they can be implemented. As a function of the number of layers and qubits, circuits following a Ring topology will have the highest entanglement and expressibility values, followed by Linear/All-to-all almost together and the Star topology. In addition to the characterization of the differences between the entanglement and expressibility of these circuits, we also place a connection between how steep is the increase on the uniformity of the distribution of the generated states and the generation of entanglement. Circuits generating average and standard deviation for entanglement closer to values obtained with the truly uniformly random ensemble of unitaries present a steeper evolution when compared to others.
\end{abstract}

\maketitle

\section{\label{sec:introduction}Introduction}

In recent times, discussions surrounding quantum advantage have become increasingly prevalent, accompanied by ongoing debates and controversies \cite{gibney2019hello,arute2019quantum,pan2022solving}. Together with the expressive increase in the investments on Quantum Technologies during the last $12$ years \cite{quantumreport_2024}, came the pressure of providing quantum computing solutions to problems of interest. It is known that the quantum algorithms and protocols that launched the Second Quantum Revolution as a great promise cannot be implemented in contemporary quantum computers \cite{second_quantum_revolution, grover_algorithm, Shor_algorithm}, due to the current limitations that avoid the near-term possibility of fault-tolerant quantum computing. Despite the persistent challenges in achieving reliable high-quality quantum error correction, the concept of noisy intermediate-scale quantum (NISQ) computers has emerged as a promising avenue \cite{preskill2018quantum}. These quantum devices exhibit potential in tackling intricate problems that may surpass the capabilities of classical computers. One possible approach in unlocking the power of NISQ computers involves the implementation of parameterized quantum circuits (PQCs). Designed with adjustable parameters, PQCs introduce flexibility into the quantum computation, which, in turn, enable an extensive exploration of the Hilbert space with a fixed architecture. When combined with classical optimizers and by defining a cost function, they give rise to variational quantum algorithms (VQAs) \cite{Cerezo2021VqaReview, peruzzo2014variational, McClean_Ucoupledcluster}. The cost function is a trainable function used in hybrid quantum-classical algorithms to perform a specific task and it might, for instance, quantify the expected value of an observable with respect to the quantum state generated by the parameterized quantum circuit, whose value is minimized or maximized to solve the problem at hand \cite{HEA_superconductors_smallmolecules, groupinvariant_QML}. Together with a classical optimizer, it is possible to minimize this cost function, thereby bringing the quantum state closer to the desired solution.

Before the actual training, however, there is the task of constructing the PQC architecture, which is also called ansatz. In many of the existing works, the PQC construction has been grounded in heuristic proposals, often drawing inspiration from established methods, such as adiabatic quantum computing and chemistry Hartree-Fock methods \cite{peruzzo2014variational,lee2018generalized,farhi2014qaoa,hadfield2019quantum,wecker2015progress,HEA_superconductors_smallmolecules, HVA_ansatz}, or imposing specific restrictions to the resulted circuit \cite{groupinvariant_QML, practicalimplementation_HEA}. In this work, rather than relying on heuristics and problem-driven inspiration for the circuit structure, we have opted to explore the potential topology of connections feasible within presently available quantum platforms. Research about the restriction on the connectivity of qubits has found relevance in various domains, being explored in the context of VQA optimization for thermodynamics protocols \cite{ivanmedina2023vqeinspired}, ground state search \cite{practicalimplementation_HEA}, and to understand the role of the entanglement \cite{PRA_Calibrating_entang_VQA}. Outside the scope of VQA, it has relevance in fault tolerant quantum algorithms \cite{circuitstopology_faulttolerant}, analysis of superconducting quantum hardware \cite{topology_superconductingqubits}, and pseudorandom quantum circuits \cite{Viola_parameters_of_pseudorandomcircuits}.

Still, PQCs have characteristics attractive even outside the borders of VQAs applications. Recent results have shown that these circuit structures can present characteristics of pseudorandom quantum circuits \cite{express_entang_capab, Kim_entanglementdiagnostics_VQA_randomcircuits,Kim_DarioRosa_chaosandcircuitparameters_randomcircuit, entanglement_production_PQC_mangini}, such as saturated entanglement, following an area or volume laws, and distribution of states that resemble an uniformly distributed measure over state space. Pseudorandom quantum circuits are quantum circuits related with unitary operators that are randomly distributed over the group manifold \cite{randomcircuits_emerson_lloyd,Viola_parameters_of_pseudorandomcircuits, violastudent_thesis}, many times compared with the uniformly distributed Haar measure. These circuits can be applied in many important quantum computing tasks, and find interesting characteristics to the generation of uniformly random unitaries \cite{randomcircuits_emerson_lloyd}, to the simulation of many-body systems dynamics \cite{randomcircuits_review}, complexity theory \cite{chaosandcomplexity_beniyoshida} and scrambling \cite{lloyd_entanglement_complexity_scrambling, Kim_DarioRosa_chaosandcircuitparameters_randomcircuit}. In this sense, to understand how this different circuit structure, the PQCs, can replicate random circuits to some extent is of fundamental and practical concern.

The enhanced exploration of the Hilbert space, facilitated by PQCs, is quantified through the concept of expressibility \cite{express_entang_capab}. It measures how uniformly distributed are the output states within the Hilbert space and it can be correlated with trainability issues in variational quantum algorithms, such as the existence of barren plateaus \cite{CerezoAkira2021BarrenPlateaus,cerezo2022diagnosing}. This relation can be best seen in Ref. \cite{holmes2022connecting} and stems from the fact that if our circuit can reach uniformly distributed states in the Hilbert space, it may be harder to reach the specific solution state we are looking for. However, a low expressible set of output states might not even reach the desired solution, so a careful analysis of this figure of merit is relevant. Also, this quantifier provides useful information from the viewpoint of the capability of PQCs to achieve pseudorandom quantum circuits, as it is measuring how close the circuit states are to a uniform random distribution over state space.

Another crucial consideration is the evaluation of how our circuit generates entanglement \cite{vedral1997quantifying,vedral1998entanglement,horodeckireview_quantument,guhne2009entanglement}. Highly entangling PQCs are undesirable in the context of VQAs, as they can lead to barren plateaus many times associated with the global distribution of local information \cite{entanglement_induced_barrenplateaus, Kim_entanglementdiagnostics_VQA_randomcircuits, Kim_DarioRosa_chaosandcircuitparameters_randomcircuit}, harming the obtainment of information through the cost function. The reason why this is happening is also associated with the convergence of the PQC to a pseudorandom circuit and $t-$designs, which will present saturated and high entanglement values \cite{scott_baker_map,lloyd_entanglement_complexity_scrambling}. Therefore, in the realm of quantum algorithms and their applications is of fundamental interest to understand the entanglement generation in quantum circuits. Referred as entangling capability \cite{express_entang_capab}, the average entanglement can be computed for the output states of random unitaries and provides information about what is the distribution of entanglement. Many different quantifiers can be chosen to address this characteristic, but an interesting one is the order $m$ Scott multipartite entanglement quantifier \cite{Scott_entanglement_measure}. This measure reduces to the well known Meyer-Wallach quantifier when $m=1$ \cite{meyer_wallach_measure, meyer_wallach_brennen, Viola_parameters_of_pseudorandomcircuits} and has an interpretation in terms of the linear entropy \cite{linear_entropy}. Both the average and standard deviation values of this function for the output states of PQCs are explored here.

In this work, we discuss the expressibility and average multipartite entanglement of parameterized quantum circuits with restrictions on the qubits connectivities. Applying the expressibility \cite{express_entang_capab} and the average entanglement for the Scott quantifier of order $1$ and $2$ \cite{Scott_entanglement_measure}, we discuss the consequences of the restrictions on the generation of quantum states. We also show that, in fact, these quantities are connected and the topology of the quantum architecture has consequences for the observed features. PQCs generating random states with average and standard deviation of entanglement closer to the observed for the ensemble of uniformly distributed random unitaries have a steeper increase in expressibility. The paper is structured as follows. In Section \ref{sec:PQC} we introduce the PQCs investigated in this work and the motivation for the choices. In Section \ref{sec:figuresofmerit} we define the quantifiers applied for the investigation, i.e., the expressibility and the average Scott quantifier. In Section \ref{sec:results}, we discuss the results for the proposed PQCs. Finally, the conclusions are presented in Sec. \ref{sec:conclusion}. {Table \ref{tab:summary} summarizes the concepts and results considered in this work.}

\afterpage{\begin{widetext}
\begin{table*}[htb]
	\caption{{Summary of quantifiers, concepts and key findings of this work.}}\label{tab:summary}%
		\centering{%
		\begin{tabular}{  l  p{12cm}}
			\toprule
			Expressibility & {A quantifier based on the Kullback-Leibler divergence, i.e., the relative entropy between the distribution of fidelities generated by a quantum circuit compared to the analytical Haar-random case. It characterizes how close is the distribution of states generated by the circuit compared to the uniform random case.}\\
			\midrule
			Average entanglement & {Considering a sample of output states, it quantifies the average value of entanglement for a particular entanglement quantifier. In this work, the generalization of the Meyer-Wallach quantifier, the Scott quantifier, with bipartition sizes $1$ and $2$ are considered.} \\
			\midrule 
			Topologies & {Different types of connectivities between qubits that can be performed in a quantum hardware. The word topology is a reference to the different graph topologies related to each of the connectivities. The chosen topologies in this work are No Connections, Linear, Ring, Star, and All-to-all.} \\
			\midrule 
			Ansätze & {For each of the topologies, many different quantum circuits can be produced depending on the gates applied. These are the so called ``ansätze''. In this work, two ansätze were considered: The first only with local rotations and a step of connections between qubits using CNOT gates. The second has the same structure of the first one, and an additional step of local rotation afterwards. They are dubbed Ansatz $1$ and Ansatz $2$, respectively.}\\
			\midrule 
			Key findings & \begin{itemize}
			    \item {Each of the circuits was studied in terms of the number of layers and qubits. With the difference topologies and ansätze, it was possible to discuss the role of the connections (not parameterized) and of the parameterized local operations;
                \item The evolutions depend on the topology, with Ring topology giving overall the highest values of expressibility and entanglement, followed by Linear/All-to-all and Star topology. However, all connected topologies lead to behaviors similar to pseudorandom circuit up to some order;
                \item The average entanglement profile influences the expressibility evolution and circuits generating mean entanglement and standard deviations closer to the random unitary matrices ensemble have a steepest expressibility increase in higher dimensions;
                \item The standard deviation of entanglement influences the freedom of the states that can be built using the circuits, which in turn influences the evolution to the uniformly random behavior;
                \item Applying the two ansätze, it was possible to isolate the effects of entanglement and we could see that the effects of additional local freedom of building coherences implies in more uniformly distributed states;
                \item We can select particular architectures based on the needs for the implementation of a quantum algorithm using the information provided by these quantifiers. For example, Star circuits are less inclined to high expressibility, and can have more potential as PQCs for VQAs, while Ring circuits presented the steepest increase of expressibility, and shall be useful as random circuits.}
			\end{itemize} \\
            \bottomrule
			\end{tabular}%
	}
\end{table*}
\end{widetext}}

\section{\label{sec:PQC}Parameterized Quantum Circuits}

\subsection{\label{subsec:circuits}Circuit Architectures and Ansätze}

\begin{figure}[htb]
    \centering
    \includegraphics[width=\columnwidth]{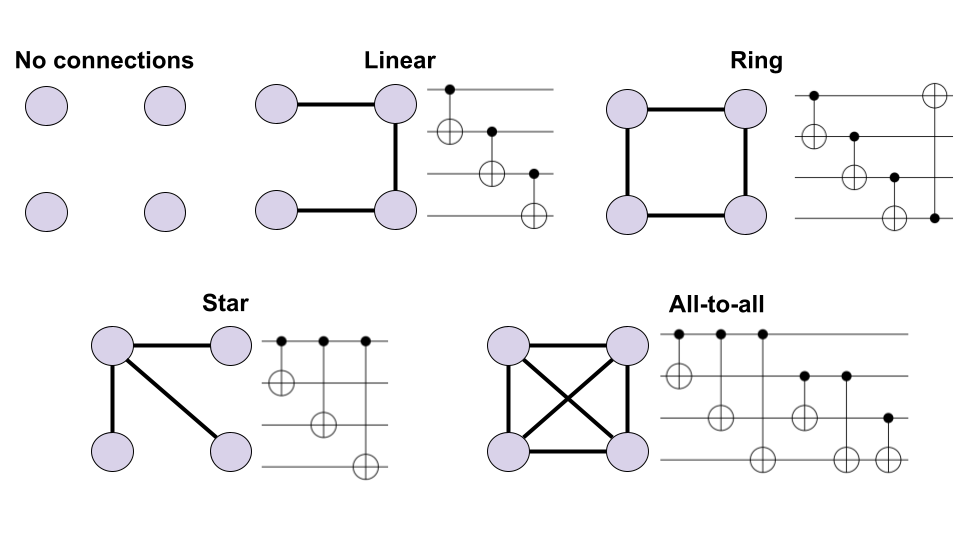}
    \caption{Graphs of the topologies listed and observed in different quantum computer architectures, together with their quantum circuit representations consisting of CNOT gates couplings.}
    \label{fig:circuits_topologies}
\end{figure}

\begin{figure}[htb]
    \centering
    \includegraphics[width=\columnwidth]{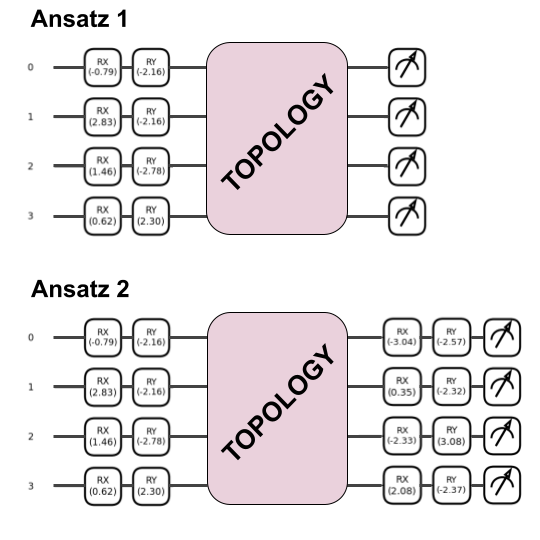}
    \caption{One layer of the circuits Ansatz $1$ and Ansatz $2$. The parameters represented in the figure were randomly sampled from a uniform distribution in the interval $[-\pi,\pi]$.}
    \label{fig:circuits_ansatze}
\end{figure}

The PQCs are one of the building blocks of a VQA and their study is central in the analysis of possible advantage and to guarantee trainability, while being constrained by the experimental reality \cite{groupinvariant_QML, entanglement_production_PQC_mangini, express_entang_capab, expressibility_new_notconfused, sauvage_spatial_symmetries_PQC}. From the perspective of random circuits applications, their implementability in real world experimental setups is also of fundamental importance \cite{Viola_parameters_of_pseudorandomcircuits, carlo_analysis_randomcircuits}. Taking this into account, in this work we consider qubits connectivities (i.e., quantum hardware topologies) that appear in contemporary and commercially available quantum hardware. Considering these different structures related with quantum computer architectures, we proposed different PQCs to understand which are the consequences of connections restrictions to the average entanglement generation and expressibility. We divided the PQCs choices in two classes, named Ansatz $1$ and Ansatz $2$, whose structure depend on the different topologies that appear in quantum computers. These different topologies are presented in Fig. \ref{fig:circuits_topologies}, while the two ansätze are shown in Fig. \ref{fig:circuits_ansatze}.

In Fig. \ref{fig:circuits_topologies}, the connectivities for $4$ qubits and their quantum circuit representations are presented. The use of CNOTs for two qubits gates is due to their nativity in many quantum computers platforms and to the fact that they are not parameterized gates. Using this strategy, we can isolate the role of the connection types, appearing in the CNOT gates, from the role of the parameterized gates, appearing in the ansätze only as local gates. We chose the $4$ qubits case to exemplify the graphs because it is the smallest situation where the connectivities are not degenerated. The simplest case in Fig. \ref{fig:circuits_topologies} is the No Connections topology (NC), as it is composed by only local qubit operations and can be implemented in any quantum computer. The availability of the other connectivities will depend on the hardware. The Linear topology (LIN) can be implemented in most of the quantum platforms, as IBM (Vigo \cite{ibm_vigo}, Tokyo \cite{topology_generativemodel_ibm} and Ithaca \cite{topology_topgen_circuitgenerator}) or Rigetti \cite{topology_generativemodeling_rigetti} quantum computers, and is composed of two qubits gates between nearest neighbours. The Ring (RIN) and Star (ST) are also available in the same platforms, however their possibility will depend on the number of qubits in the circuits. The Ring connects first neighbours and the first to the last qubit in a circuit, while Star has connections between one central qubit and every other qubit. The most complex topology, All-to-all (ATA), will have connections of every qubit with every qubit (a complete graph) and can be implemented in ion traps quantum computers with a high number of qubits \cite{topology_iontraps_alltoall} without the need of additional SWAP gates. When considering $3$ qubits, the smallest dimension studied here, the Linear/Star and Ring/All-to-all topologies will be degenerated, therefore we are going to dub the two equivalent sets as Linear and Ring.

Now, these topologies will be part of the circuit structure presented by ansätze $1$ and $2$, Fig. \ref{fig:circuits_ansatze}, applied to an initial state $\ket{0}^{\otimes n}$. The local parameterized rotations used in the ansätze are given by ${\rm Rj}(\theta)=e^{-i \theta \sigma_j}$, where $j=$X,Y, $\sigma_j$ are the usual Pauli matrices and $\theta$ is an arbitrary parameter representing the rotations angles. The Ansatz 1 is composed of sequential applications of parameterized rotations RX and RY on each qubit, followed by the connections according to the chosen topology. The Ansatz 2 shares this same structure with an additional application of parameterized rotations RX and RY after the connections step. Both circuits considering only $1$ layer will share the same average entanglement, as the operations after the connections in Ansatz 2 are only local unitary operations that preserve entanglement values. We will show that, in fact, this will be also true for more than $1$ layer. From this characteristic, we can affirm that Ansatz 2 will have the same entanglement as Ansatz 1, however with more freedom on which are the local coherences the different states can have. It is worth remarking that the parameters are encoded only in the local unitary rotations. Therefore, we can separate the roles played by the topology, defined only by the connective part of the circuits (CNOTS gates), and by the local operations, which imprint the parameters in the system state, leading to local superposition. This will be of fundamental importance to understand our results and how these circuits can be useful in different protocols. 

We chose local operations RX and RY due to the fact that they are going to generate random $1$ qubit states over the complete surface of the Bloch sphere sampling the circuit parameters uniformly, when starting with a $\ket{0}$ state. One of the drawbacks of this choice is that the random $1$ qubit states will present a bias towards two of the axis of the sphere, however the states generated effectively cover the space \cite{express_entang_capab}. {There are other possible approaches to explore the $1-$qubit Hilbert space uniformly, however they would require different probability distributions or gate configurations. Here, we chose to follow the choice provided by the original paper where the expressibility and entangling capability were first introduced, considering the uniform distribution in a fixed interval to not infer any bias over the parameters, and RX and RY gates to be faithful to our goal of applying parameterized rotations.} Table \ref{tab:numberofgates_ansatze_top} presents the number of CNOT gates and total number of gates in both circuits Ansatz $1$ and $2$. Comparing the different topologies, as a function of the number of qubits, $n$, and number of layers, $l$. The total number of parameters appears isolated inside brackets in the equations of the total number of gates, so this number can be easily known looking at the second term inside brackets. For example, for the Linear topology and Ansatz $2$, there are $\left[(n-1)+4n\right]l$ gates, being $(n-1)l$ CNOTs and $(4n)l$ parameterized gates.

\begin{table*}[htb]
	\label{tabela}
	\caption{Number of CNOTs and total number of gates comparing topologies for Ansatz $1$ and $2$ as a function of the number of qubits $n$ and number of layers $l$. The terms inside brackets are separated so the right term in each of them is exactly the number of parameters (total number of gates RX and RY) in each circuit.}\label{tab:numberofgates_ansatze_top}%
		\centering{%
		\begin{tabular}{cccc}
			\toprule
			Topology & Number of CNOTs & Total number Ansatz $1$ & Total number Ansatz $2$\\
			\midrule \midrule
			No Connections & $0$ & $(2n)l$ & $(4n)l$\\
			\midrule 
			Linear & $(n-1)l$ & $\left[(n-1)+2n\right]l$ & $\left[(n-1)+4n\right]l$\\
			\midrule 
			Ring & $nl$ & $\left[n+2n\right]l$ & $\left[n+4n\right]l$\\
			\midrule 
			Star & $(n-1)l$ & $\left[(n-1)+2n\right]l$ & $\left[(n-1)+4n\right]l$\\
            \midrule
            All-to-all & $\frac{n(n-1)}{2}l$ & $\left[\frac{n(n-1)}{2}+2n\right]l$ & $\left[\frac{n(n-1)}{2}+4n\right]l$ \\
            \bottomrule
			\end{tabular}%
	}
\end{table*}

\section{\label{sec:figuresofmerit}Figures of Merit}

\subsection{\label{subsec:expressibility}Expressibility}

In this work we are concerned about the capability of PQCs to generate uniformly distributed random states over the Hilbert space of interest. When considering random pure quantum states in a Hilbert space of any finite dimension $d=2^n$, being $n$ the number of qubits, a uniform probability measure $d\mu(\psi)$ can be obtained using the Fubini-Study (FS) metric \cite{MetricinducedEnsembles_Hall1998}. This same measure can be also obtained by a different method. Considering the uniform probability measure over the unitary group of dimension $d$, $\mathcal{U}(d)$, sample unitaries and act with them over a fiducial state, here defined as the tensor product $\ket{0}^{\bigotimes n}$. The  obtained ensemble of states will have the same distribution as the one sampled using the FS-metrics over the space of states \cite{how_to_generate_rmatrices_mezzadri, Measuresspaceofstates_Zyczkowski_2001}.

This measure over the unitary group is called Haar measure and presents interesting properties as invariance considering elements of the group \cite{haarmeasure_generalintroduction}. However, building a faithful replication of the Haar measure can be hard and costly, requiring circuits with gate complexity that scales exponentially with the dimension of the system \cite{randomcircuits_emerson_lloyd}. Also, only part of the properties of the Haar measure are needed to determine integrals of polynomials of restricted order or mean values of operators, motivating the definition of $t-$designs \cite{unitary_designs, polynomial_t-designs1, polynomial_t-designs2}. The $t-$designs are ensembles of unitaries that can replicate parts of the characteristics of the Haar measure over the unitary group, with different plausible equivalent definitions \cite{haarmeasure_generalintroduction}. One possible definition is in terms of the polynomials one might want to calculate integrals over the group manifold.

\begin{defin}\label{def:unitary_design_polynomials}
\textbf{(Unitary $t-$design via polynomials)}\cite{polynomial_t-designs1, polynomial_t-designs2} Let $\{U(x)\}_{x=1}^K\subset\mathcal{U}(d)$ be a finite set of unitary operators acting on the space $\mathbb{C}^d$. If the set is such that for every polynomial $P_{(t,t)}(U)$ of degree at most $t$ in the elements and complex conjugate elements of the matrix $U$,

    \begin{equation}
        \frac{1}{K}\sum_{x=1}^K P_{(t,t)}(U(x)) = \int_{\mathcal{U}(d)}d\mu(x)P_{(t,t)}(U(x)),
    \end{equation}
    \noindent
    is satisfied, we say the set is a unitary $t-$design. The $d\mu(x)$ is the Haar measure over $\mathcal{U}(d)$.
\end{defin}

A measure to compare true $t-$designs with a distribution of states induced considering a parameterized circuit structure sampling the parameters was introduced in Ref. \cite{express_entang_capab}. It is defined in terms of the square of the Hilbert-Schmidt norm of the quantity

\begin{equation}
    \mathcal{A}^t := \int_{\text{Haar}}(\ket{\psi}\bra{\psi})^{\otimes t}d\ket{\psi} -
    \int_{\Theta}(\ket{\phi_\theta}\bra{\phi_\theta})^{\otimes t}d\theta,
\end{equation}
that compares the induced Haar state a $t-$design would construct (first integral) \cite{chaosandcomplexity_beniyoshida} with the one induced when considering the circuit and sampling over the parameter space (second integral). The sampling in this work is performed with a uniform probability measure between $0$ and $2\pi$ for each individual parameter. This norm can be translated in terms of the statistical moments of fidelities comparing states generated considering the Haar ensemble and the circuit ensemble \cite{express_entang_capab},

\begin{equation}\label{eq:t_design_quantifier}
||\mathcal{A}^t||^2_{\text{HS}}:=\Tr\left[(\mathcal{A}^t)^\dagger(\mathcal{A}^t)\right]=\mathbb{E}\left[F(\boldsymbol{\theta},\boldsymbol{\phi})^t\right]-\mathbb{E}_{\text{Haar}}(F^t),
\end{equation}
being $F(\boldsymbol{\theta},\boldsymbol{\phi}):=|\braket{\psi(\boldsymbol{\theta})}{\psi(\boldsymbol{\phi})}|^2$ the fidelity \cite{chapter_Diogo_geometricquantumcorrelations} for two circuit generated pure states and the second expected value obtained by the calculation of fidelities comparing two Haar random states. This will be greater or equal to zero and the closer it is to zero, the closer the circuit structure is to generate a $t-$design.

This quantifier considers a particular $t-$design order. To compare with the Haar measure in a general form, the expressibility quantifier was introduced \cite{express_entang_capab} based on the Kullback-Leibler divergence or relative entropy \cite{wilde_shannontheory} between the circuit distribution of fidelities and the Haar one, whose analytical values are known for pure states in finite Hilbert spaces \cite{fidelitiesprobability_zyczkowski}. The expressibility is then defined as \cite{express_entang_capab}

\begin{equation}
    \text{Expr}:=D_{KL}\left(P_{PQC}(F)||P_{\text{Haar}}(F)\right),
\end{equation}
where $P_{PQC}$ is the circuit fidelity histogram and $P_{\text{Haar}}$ the Haar one. The closer this quantity is to zero, the closer the circuit induced ensemble is to generate uniformly distributed states in the space of states and we say the more expressible is the circuit.

To compute this quantifier, the parameter vectors are sampled and given as input for a particular circuit structure. A output state is built for a sampled parameter vector and the input $\ket{0}^{\otimes n}$ and with two of these states, is possible to sample a fidelity value. We consider $10^4$ parameter vectors according to the Chebyshev inequality \cite{Ross_firstcourse_in_probability}, giving rise to $5\cdot 10^3$ fidelities that will compose the circuit histogram to be compared with the Haar histogram that can be analytically built.

\subsection{\label{subsec:scottentanglement}Scott entanglement measure and average entanglement}

The Scott entanglement measure was proposed to quantify multipartite entanglement considering the mean bipartite entanglement over every possible bipartition of a particular size \cite{Scott_entanglement_measure, Rigolin_entanglement_measure2}. It was introduced based on the decomposition of the Meyer-Wallach entanglement measure (MW) \cite{meyer_wallach_measure, meyer_wallach_brennen} in terms of the linear entropy. The MW computes multipartite entanglement based on the mean linear entropy \cite{linear_entropy} over every possible bipartition with size $1$ qubit-rest of the system. Scott generalized this to any bipartition of size $m$ qubits and the rest of the system. Given an $n$ qubits pure state, $\rho(n)\equiv\ketbra{\psi}{\psi}$, the Scott entanglement measure of order $m$ \cite{Scott_entanglement_measure}, that we are going to refer either as $Q_m$ or $S_m$, reads

\begin{eqnarray}
    Q_m\left(\ket{\psi}\right)&:=&\frac{2^m}{2^m-1}\left[1-\binom{n}{m}^{-1}\sum_{|S|=m}\Tr\left(\rho_S^2(n)\right)\right], \nonumber\\
    m&\leq& \lfloor n/2\rfloor,
\end{eqnarray}
where $S$ is a size $m$ subset of $\{1,2,\cdots,n\}$ indexing the qubits in the system, $\rho_S(n)$ is the reduced state discarding the complementary set $S^{'}$ such that $S\cup S^{'}=\{1,2,\cdots,n\}$ and $\lfloor\cdot\rfloor$ is the floor function. This way, the order $m$ defines the sizes of the bipartitions considered. The constraint over $m$ defines the normalization to be always valid and avoid redundancies for different orders. For example, if $n=3$, $m=1$ and $m=2$ give the same information about the multipartite entanglement. When $m=1$, the bipartition size considered will be $1$ qubit and the rest of the system, that is composed of $2$ qubits. The linear entropies of reduced states with this bipartition are completely equivalent to the ones considering $m=2$, where the parts are $2$ qubits and the rest of the system, i.e., $1$ qubit. Therefore, both $m=1$ or $m=2$ represent the same entanglement measure.

The different values of $m$ in the Scott entanglement measure will explore different characteristics of multipartite entanglement and will define different entanglement measures. For example, there are states for which the MW, $m=1$, will present the same entanglement values, even though they have very different characteristics. We can cite the states of $6$ qubits, {$\ket{GHZ_6}=(\ket{0}^{\otimes 6}+\ket{1}^{\otimes 6})/\sqrt{2}$} and $\ket{EPR_6}=\ket{\Phi^+}^{\otimes 3}$, being $\ket{\Phi^+}=(\ket{00}+\ket{11})/\sqrt{2}$, that will have $Q_1(\ket{GHZ_6})=Q_1(\ket{EPR_6})=1$. These states have very different entanglement characteristics, being the $\ket{GHZ_6}$ sometimes considered a state with maximum multipartite entanglement, while the $\ket{EPR_6}$ is a type of biseparable state \cite{wyderka_thesis, threequbits_entanglementclasses}. Choosing $m=2$, this degeneracy is broken and we have $Q_2(\ket{GHZ_6})=2/3$ and $Q_2(\ket{EPR_6})=4/5$. Therefore, in this work we compute the average entanglement considering the measures $S_1$ and $S_2$, so we can understand if the results can be replicated with both entanglement measures.

To quantify the behavior of the generation of entanglement in the circuits, we applied a method very similar to the expressibility computation one \cite{express_entang_capab}. We generate random parameter vectors considering the uniform distribution and, given the input $\ket{0}^{\otimes n}$ state, calculate the entanglement for the output states and obtain the average and standard deviation over the different parameters,

\begin{eqnarray}
    \langle Q_m \rangle_{\Theta}&=&\frac{1}{\mathcal{N}}\sum_{i=1}^{\mathcal{N}}Q_m(\ket{\psi(\boldsymbol{\theta}_i)}), \nonumber \\
    \sigma_{\Theta}(Q_m) &=& \sqrt{\langle Q_m^2\rangle_{\Theta}-\langle Q_m\rangle_{\Theta}^2}.
\end{eqnarray}
In previous works, only the average entanglement was considered \cite{express_entang_capab, entanglement_production_PQC_mangini} for a thorough analysis. We can compare these numerical calculations with the analytical values considering uniformly distributed random states, obtained with the Haar induced distribution, named Circular Unitary Ensemble (CUE). The values for the averaged $S_m$ and standard deviation for $S_1$ are \cite{Scott_entanglement_measure, scott_baker_map}

\begin{widetext}
\begin{eqnarray}\label{eq:analyticalaverage_CUE}
    \langle Q_m \rangle_{CUE} &=& \frac{2^n-2^m}{2^n+1}, \nonumber \\
    \sigma_{CUE}(Q_1) &=& \sqrt{\frac{6(2^n-4)}{(2^n+3)(2^n+2)(2^n+1)n} + \frac{18\cdot2^n}{(2^n+3)(2^n+2)(2^n+1)^2}}.
\end{eqnarray}
\end{widetext}

This comparison will be interesting to understand if and when it is possible to replicate the uniform random entanglement applying PQCs. From the perspective of $t-$designs, the Scott measures are polynomials of order $(2,2)$ \cite{Viola_parameters_of_pseudorandomcircuits}, so $2-$designs are enough to replicate the exact analytical average values in Eq. \eqref{eq:analyticalaverage_CUE}. When discussing the relation between the entanglement generation and expressibility changes as a function of the number of layers, this characteristic will be important.

\section{\label{sec:results}Results}

In this section we are going to discuss the values and relations between the expressibility and entanglement quantifiers. They are going to be compared considering the two parameterized structures, ansätze $1$ and $2$, with the different connectivities, for different numbers of layers and qubits. We varied the number of circuit layers between $1$ and $5$ and the number of qubits from $3$ to $8$. Exploring different number of circuit layers is interesting from the viewpoint of VQAs, where this is sometimes applied to verify performance changes and supply additional parameters for optimization \cite{Cerezo2021VqaReview, HVA_ansatz}. It is also interesting from the perspective of pseudorandom quantum circuits, where the sequential application of a particular structure evolves to a simulation of a Haar random unitary \cite{randomcircuits_emerson_lloyd, Viola_parameters_of_pseudorandomcircuits, randomcircuits_review}. To treat different dimensions is also of fundamental importance to understand how the size of the system considered will affect the results and how well the circuit will perform in these different settings. We start with the expressibility for Ansatz $1$, followed by the average entanglement. After that, we show that the entanglement is the same for both ansätze and connect with the expressibility of Ansatz $2$. The calculations were performed using the quantum circuits simulation Pennylane Python library \cite{bergholm2022pennylane} and the codes are available on github \footnote{Available on the profile https://github.com/GICorrer.}.

\subsection{\label{subsec:ansatz1}Ansatz 1}

\subsubsection{\label{subsubsec:expressA1}Expressibility}

\begin{figure}[htb]
    \centering
    \includegraphics[width=\columnwidth]{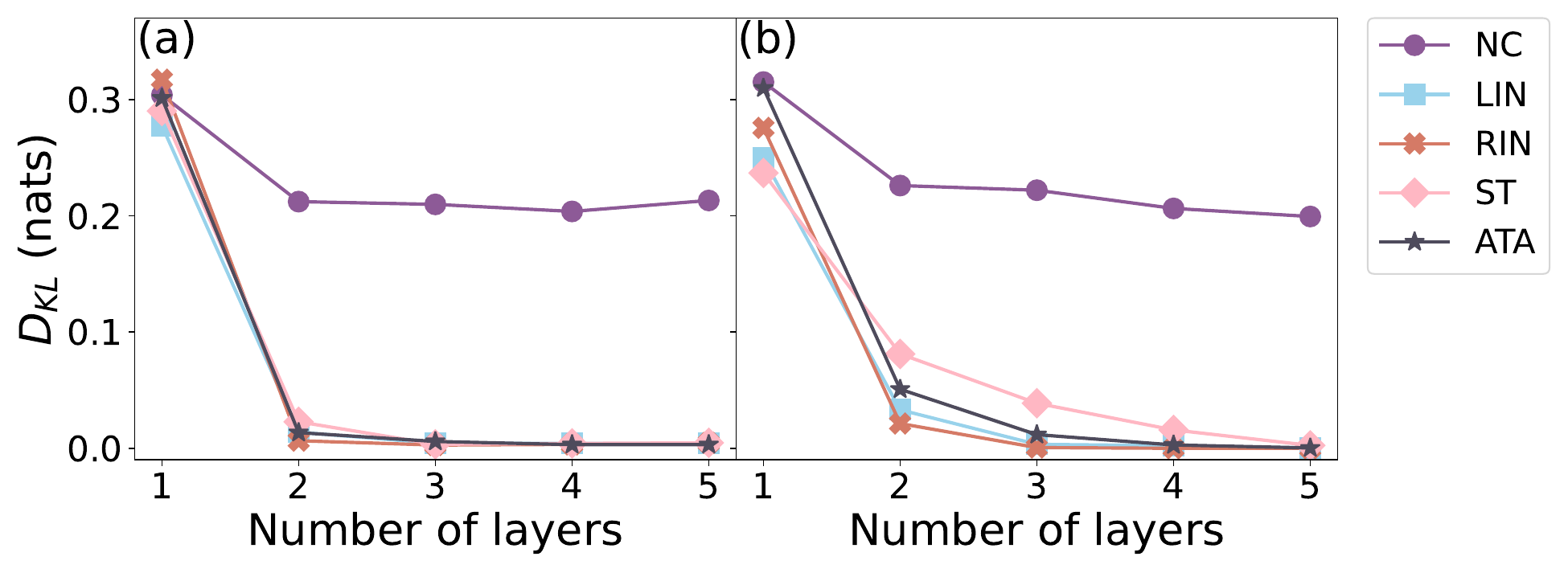}
    \caption{Ansatz 1: relative entropy as a function of the number of layers for all connectivities, considering (a) 4 qubits and (b) 8 qubits. NC: No Connections, LIN: Linear, RIN: Ring, ST: Star and ATA: All-to-all.}
    \label{fig:expressibility_ans1_4-8_qubits}
\end{figure}

In Fig. \ref{fig:expressibility_ans1_4-8_qubits}, the relative entropy as a function of the number of layers for Ansatz $1$ is shown for every connectivity with $4$ and $8$ qubits. The first observation we can make is that the No Connections circuit saturates between $0.20$ and $0.25$ nats for every studied dimension and cannot evolve to values closer to $0$. This is expected, as the input state is separable, $\ket{0}^{\otimes n}$, and the circuit does not generate entanglement. Random uniformly distributed states should have nonzero entanglement (Eq. \eqref{eq:analyticalaverage_CUE}) and if it is not generated, the circuit expressibility will saturate at a value not that close to this Haar case. Conversely, the connected topologies will evolve to values very close to $0$, indicating the increase in how expressible the circuits are. This aspect highlights that PQCs can replicate the behavior of pseudorandom quantum circuits up to some order when a sampling of the parameters is performed. The interesting feature is that PQCs will have a fixed structure that is repeated, contrasting with many proposals of random circuits whose positions of gates in the circuits are also random \cite{brandao_randomcircuitmodel, harrow_randomcircuitmodel}. Other works studied how the uniformly distributed randomness of the states generated with the circuit will affect optimization and performance in a VQA scenario \cite{express_entang_capab, Kim_entanglementdiagnostics_VQA_randomcircuits, Kim_DarioRosa_chaosandcircuitparameters_randomcircuit}, however not much attention is given to the capability of these circuits to generate random states. This behavior is observed for every dimension studied. A peculiar result is presented for only $1$ layer. Comparing the different circuit structures there is no established advantage of generating entanglement, i.e., the connected topologies present relative entropy values that are very close to the observed for the No Connections case, even though this circuit is not generating entanglement.

\begin{figure}[htb]
    \centering
    \includegraphics[width=\columnwidth]{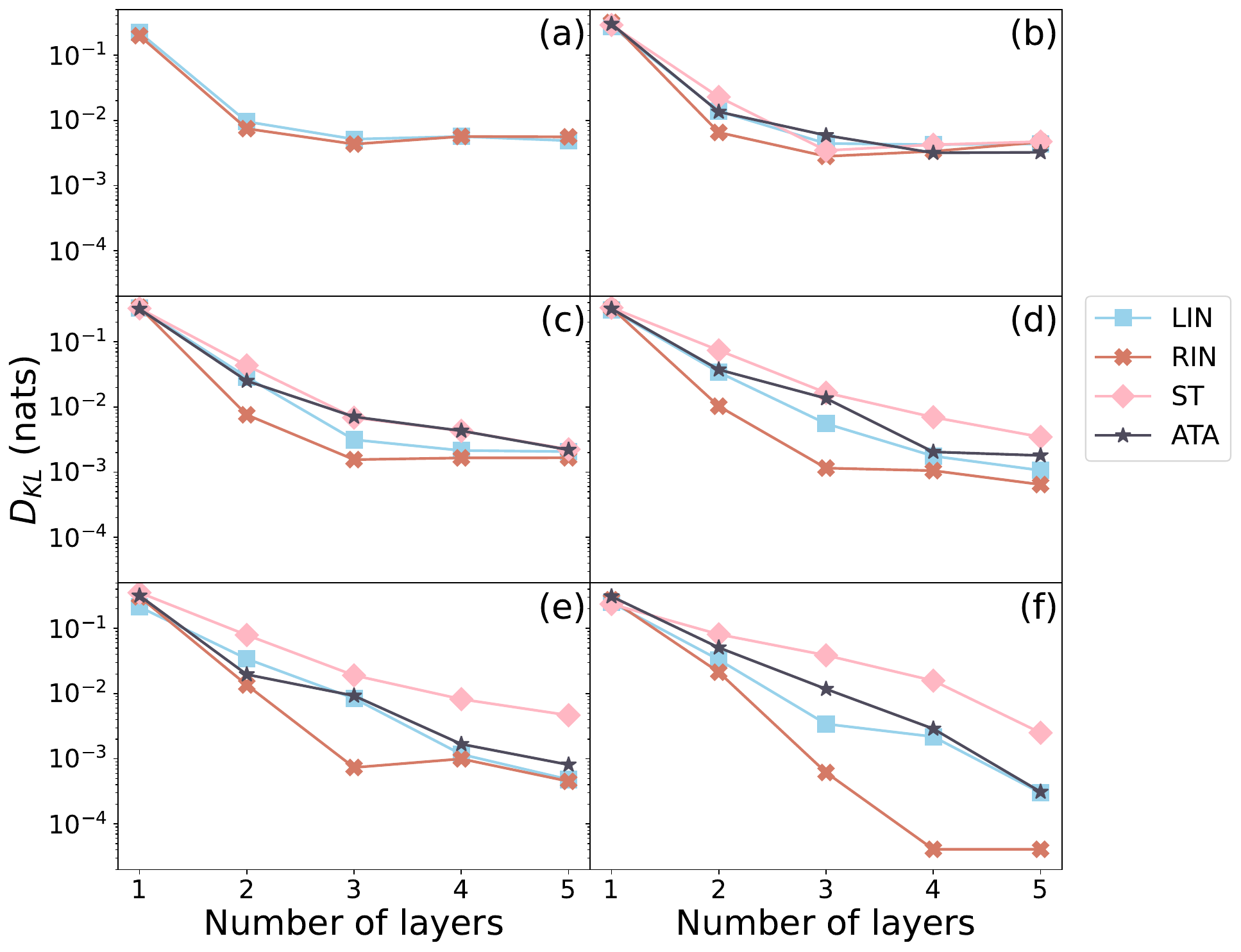}
    \caption{Ansatz 1: relative entropy as a function of the number of layers for the connected topologies, considering (a) 3 qubits, (b) 4 qubits, (c) 5 qubits, (d) 6 qubits, (e) 7 qubits and (f) 8 qubits. LIN: Linear, RIN: Ring, ST: Star and ATA: All-to-all.}
    \label{fig:expressibility_ans1_logscale}
\end{figure}

To understand how the connected topologies are different in terms of the relative entropy as a function of the number of layers, in Fig. \ref{fig:expressibility_ans1_logscale} a plot with logarithmic scale in the $y$-axis is shown. For the smaller dimensions of $3$ and $4$ qubits, Fig. \ref{fig:expressibility_ans1_logscale} (a) and (b), there is not much difference in the evolution and they follow a very similar behavior. A saturation is also observed at $5\cdot 10^{-3}$ nats for both cases. We associate this with the fact that the rotations $RX$ and $RY$ together do not define a universal set of unitaries for one qubit, a requirement for the evolution of a circuit structure to a pseudorandom circuit \cite{randomcircuits_emerson_lloyd, Viola_parameters_of_pseudorandomcircuits}. A circuit involving universal one qubit unitaries and CNOTs should converge to the generation of pseudorandom states with sufficient number of concatenations of the circuit structure \cite{randomcircuits_emerson_lloyd}. However, another plausible explanation is that the rotations $RX$ followed by $RY$ with parameters uniformly sampled are generating biased states of one qubit, with concentrations around the poles of the Bloch sphere \cite{express_entang_capab}. A more interesting approach would be to consider a sampling that replicates the uniform distribution of points in the shell of a $2-$sphere, where the angles take into account the integration measure $\sin(\theta) d\theta d\phi$, being $\theta$ the polar angle and $\phi$ the azimuthal angle. This way, the sampling of states would replicate the action of a uniformly distributed unitary.

When the dimension increases, from panels (c) to (f) of Fig. \ref{fig:expressibility_ans1_logscale}, it is possible to notice that the evolutions start to branch, establishing a hierarchy between the topologies. This difference grows with the dimension and presents the biggest differences at $8$ qubits, as it is shown in panel (f). The Star circuit presents the slowest decay to the Haar random case, followed by Linear/All-to-all almost together and finally the Ring, with the steepest decrease. Still, a saturation is observed also for $5$ qubits and apparently has begun for $6$ qubits. In smaller dimensions, the differences in the connectivities are not very influential, however in higher dimensions we can clearly observe that the way qubits can be connected affects the expressibility. In the circuits of this work, the topology only enters the problem in the CNOT gates part of the circuits, which is also responsible for the entanglement generation. In this sense, it is expected that the generation of entanglement and the expressibility of the circuits might be related since both quantities strongly depends on the topology.

\begin{figure}[htb]
    \centering
    \includegraphics[width=\columnwidth]{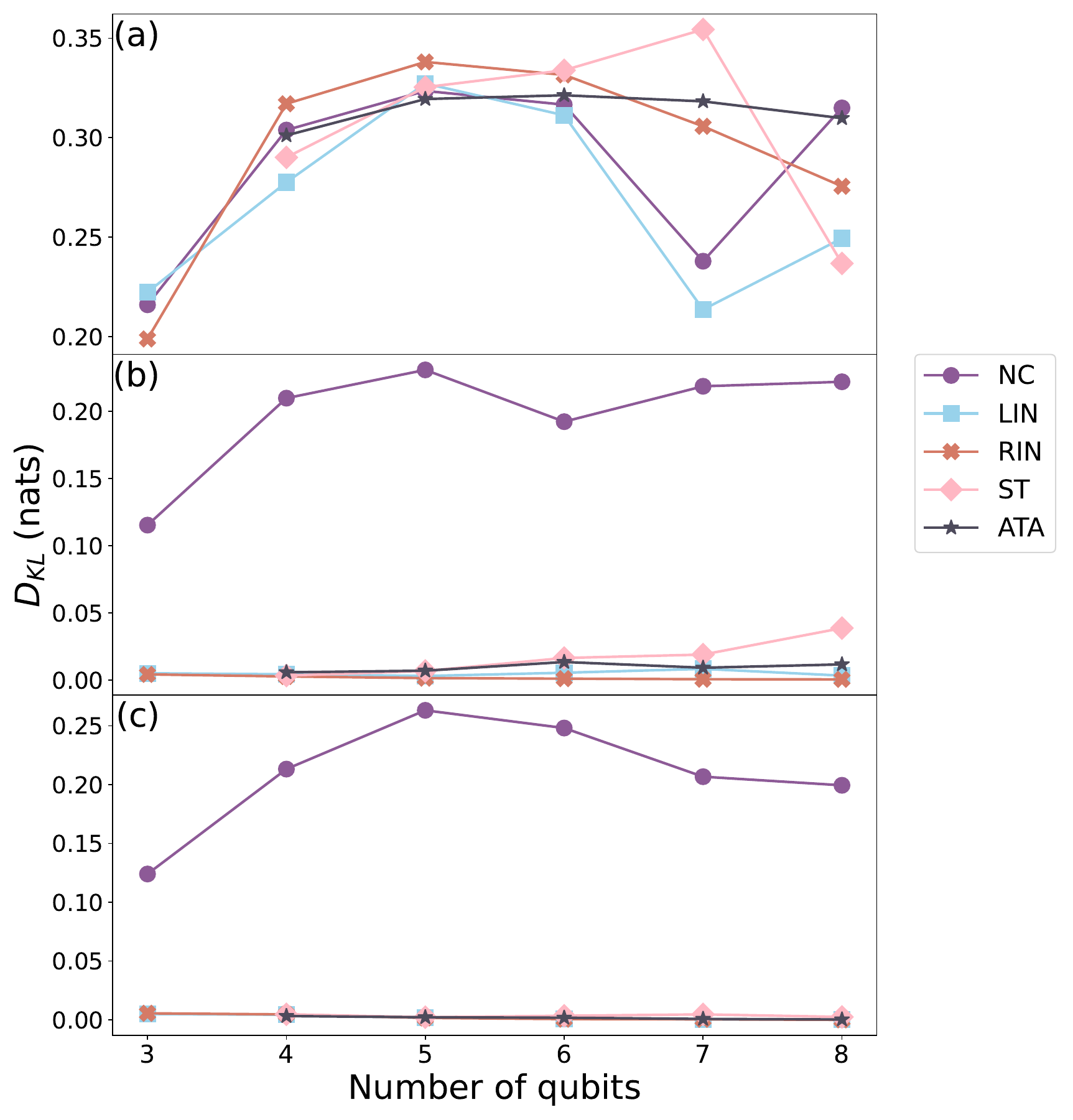}
    \caption{Ansatz 1: relative entropy as a function of the number of qubits for the connected topologies, considering (a) 1 layer, (b) 3 layers and (c) 5 layers. NC: No Connections, LIN: Linear, RIN: Ring, ST: Star and ATA: All-to-all.}
    \label{fig:expressibility_ans1_numberofqubits}
\end{figure}

Fig. \ref{fig:expressibility_ans1_numberofqubits} highlights the differences discussed in the paragraphs above using plots with fixed number of layers as a function of the number of qubits. In Fig. \ref{fig:expressibility_ans1_numberofqubits} (a), it is possible to observe that all the topologies with the circuit structure of Ansatz 1 present close values for the expressibility if we consider only $1$ layer. This is observed for every dimension. Increasing the number of layers to $3$, Fig. \ref{fig:expressibility_ans1_numberofqubits} (b), the connected topologies relative entropies decay to values closer to zero, while the not connected saturates in higher values. Increasing the dimension, we can observe the hierarchy establishment. At $5$ layers, Fig. \ref{fig:expressibility_ans1_numberofqubits} (c), the connected topologies get very close to zero, despite their differences.

\subsubsection{\label{subsubsec:entangA1}Average entanglement}

\begin{figure}[htb]
    \centering
    \includegraphics[width=\columnwidth]{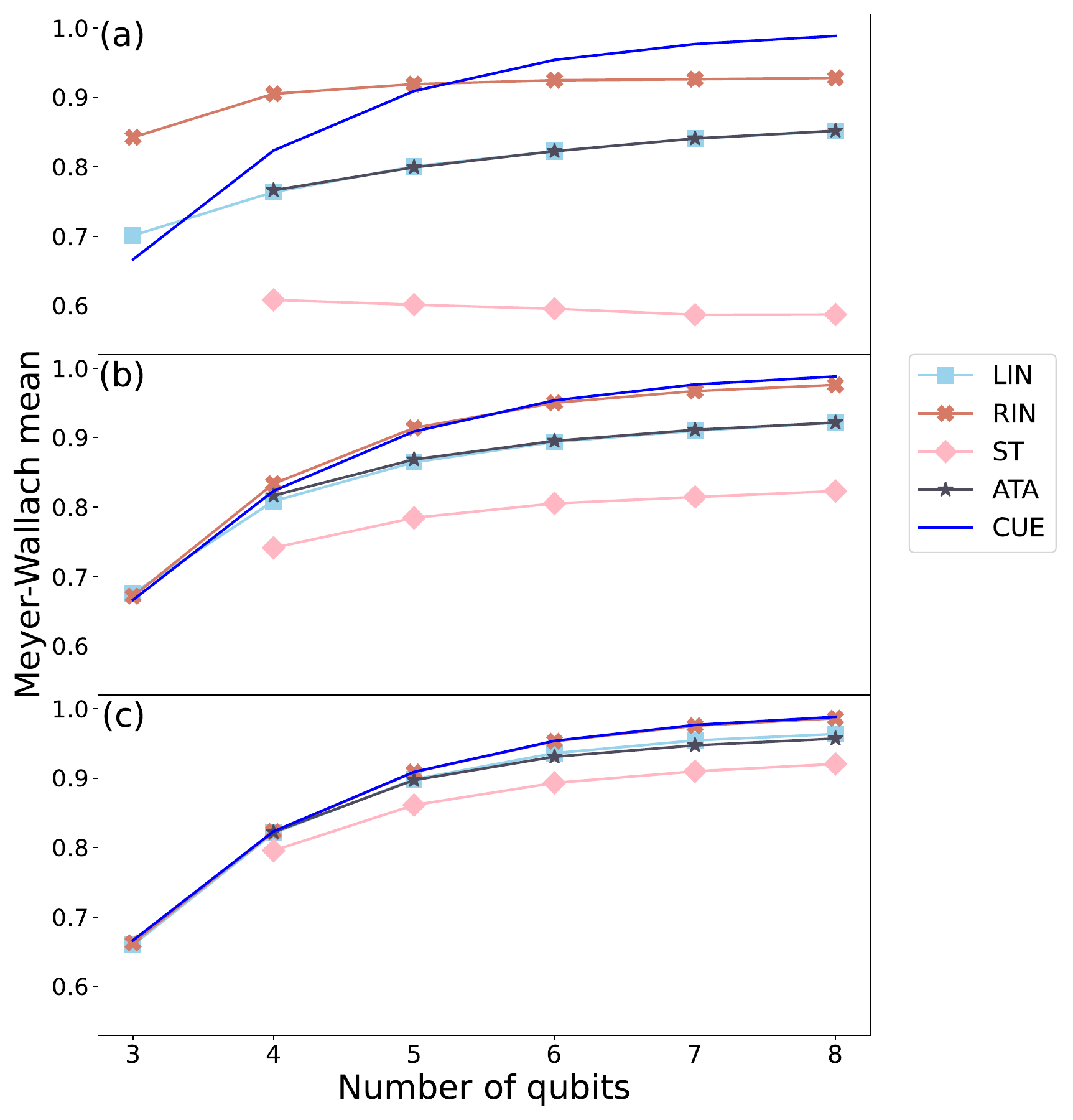}
    \caption{Ansatz 1: average Meyer-Wallach/Scott $1$ for all connected topologies as a function of the number of qubits for (a) 1 layer, (b) 3 layers and (c) 5 layers. The blue line indicates the analytical values of the CUE given in Eq. \eqref{eq:analyticalaverage_CUE}. LIN: Linear, RIN: Ring, ST: Star and ATA: All-to-all.}
    \label{fig:entanglement_ans1_numberofqubits}
\end{figure}

We begin the average entanglement results presentation by discussing the absolute values observed for the circuits. To do so,  in Fig. \ref{fig:entanglement_ans1_numberofqubits} it is shown the mean entanglement for a fixed number of layers as a function of the number of qubits. The dark blue line indicates the  analytical behavior observed for uniformly distributed states, which increases to $1$ as a function of the dimension. This way, for high dimensional systems, the average entanglement of uniformly distributed states will be very close to maximum \cite{genericentanglement_concentration_measure, scott_baker_map, Scott_entanglement_measure}. We can see that, overall, the Ring circuit generates the highest entanglement values, followed by Linear/All-to-all exactly together and by the Star with the least values. Increasing the number of layers, the circuits entanglement gets closer to the analytical values, an expected result as they generate states more uniformly distributed. In fact, we can say they are approximating a $2-$design (Def. \ref{def:unitary_design_polynomials}) when the analytical values are achieved \footnote{This affirmation has to be made with much caution, because not every circuit that generates entanglement values close to the analytical CUE values approximates a $2-$design. Still, in this case this is true as the circuits are converging to the analytical case and a relative entropy decrease was observed.}. The Ring is the only one that can achieve this reproduction with $5$ layers (Fig. \ref{fig:entanglement_ans1_numberofqubits} (c)), while the others would need more layers to faithfully reproduce it.

If we look at the dependency of the entanglement as a function of the dimension for only $1$ layer, it is possible to notice a slight increase for the Ring, All-to-all and Linear topologies, while the Star circuit preserve the same entanglement values. This can be explained by the fact that the connections structure for the Star circuit prepares entangled states that have a strong dependence on the state built with the RX-RY gates step for the central qubit. Every other qubit is connected to this one and not connected to any other, so if the state prepared for it does not favor entanglement generation with the CNOT gates step, increasing the number of qubits will not influence the entanglement generation. Therefore, the generation of entanglement is highly connected to this central qubit and changing the number of qubits, consequently increasing the number of CNOTs, will not affect entanglement values. In contrast, the generation of entangled states and distribution of CNOT gates is not confined in the other connected topologies: The connections are distributed among the qubits that compose the circuit. Therefore, we can say that increasing the number of qubits will imply in more possibilities of connections between qubits and favor the average entanglement increase with the dimension.

\begin{figure}[htb]
    \centering
    \includegraphics[width=\columnwidth]{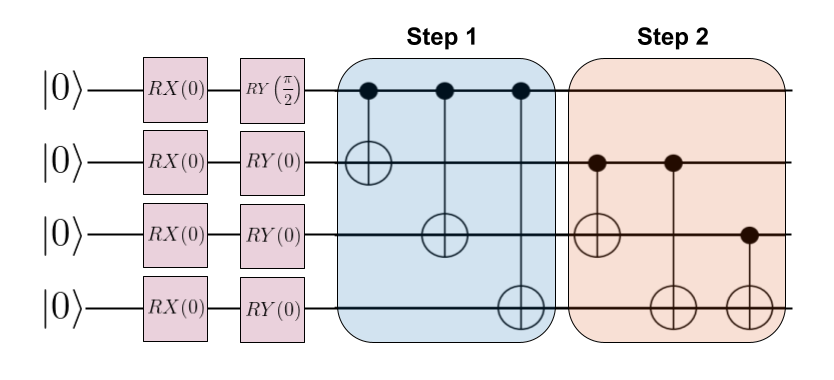}
    \caption{Specific example of a sequence of gates applied considering a $4$ qubits All-to-all circuit of Ansatz $1$. In this set, a Hadamard is applied to the first qubit and nothing is done to the other qubits in the parameterized operations step. After that, the connections are applied divided into two intermediate steps, Step $1$ (blue) and Step $2$ (red).}
    \label{fig:entanglement_destruction_ATA_GHZ}
\end{figure}

An unexpected feature of the chosen connectivities is the coincidence in the entanglement of the Linear and All-to-all topologies, even though they have very different number of CNOT gates and complexity of connections, with the All-to-all presenting a quadratic scale of number of gates with the number of qubits (see Table \ref{tab:numberofgates_ansatze_top}). To understand why, a specific example is considered. If we take one layer of a Ansatz $1$ All-to-all circuit with $4$ qubits and perform the sequence of operations defined in Fig. \ref{fig:entanglement_destruction_ATA_GHZ}, the result will be equivalent to a Hadamard gate in the first qubit, nothing in the others and the steps of connections after that. Using these gates, after the Step $1$ in Fig. \ref{fig:entanglement_destruction_ATA_GHZ}, the resulting state will be a $4$ qubits GHZ state, that has maximum entanglement quantified by the Meyer-Wallach. However, performing Step $2$, the output state will be $\ket{\psi_{\text{out}}}=(\ket{00}+\ket{11})/\sqrt{2}\otimes\ket{0}\otimes\ket{0}$, whose entanglement quantified by the Meyer-Wallach/Scott $1$ is equal to $1/2$. This way, the step of CNOT gates in the All-to-all topology is creating and sometimes destroying entanglement at the same time, influencing the average entanglement. We can then say that different parameter vectors will be able to generate highly and slightly entangled states due to the connections choice, resulting in an average that replicates the entanglement generated by the Linear connected circuit with a way more complex structure. {From another perspective, it was shown in Ref. \cite{entanglement_production_PQC_mangini} that connections following the All-to-all pattern shown in this work are trivially equivalent to a circuit with CNOTs following the pattern of a Linear (first-neighbours) graph, justifying the observed behavior that the All-to-all circuit in our work produces the exact same behavior when compared to the Linear. This can be achieved by identities and equivalences between sequence of CNOT operations \cite{equivalentquantumcircuits}.} Therefore, the choice of performing all the CNOT gates in a row is not optimized for generating high average entanglement and other choices could be more interesting. One plausible option could be to interleave the local operations with the CNOT gates, avoiding the sequential CNOTs with the same local coherences.

\begin{figure}[htb]
    \centering
    \includegraphics[width=\columnwidth]{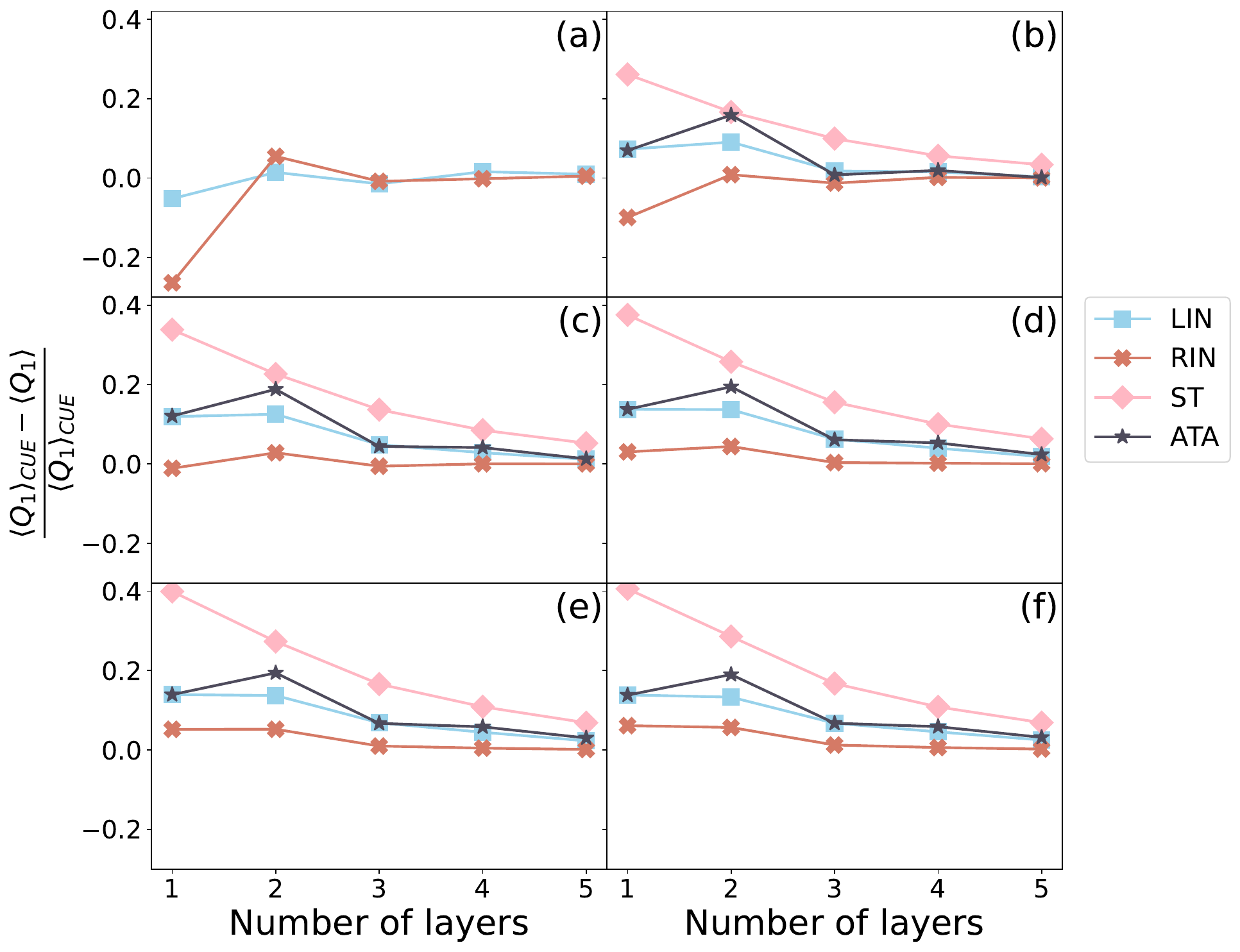}
    \caption{Ansatz 1: subtraction of the CUE ensemble mean and the circuit Meyer-Wallach/Scott 1 average, normalized with the CUE value for all connected topologies as a function of the number of layers for (a) 3 qubits, (b) 4 qubits, (c) 5 qubits, (d) 6 qubits, (e) 7 qubits and (f) 8 qubits. LIN: Linear, RIN: Ring, ST: Star and ATA: All-to-all.}
    \label{fig:entanglement_subtraction_ansatz1}
\end{figure}

As said before and demonstrated now with data, the generation of entanglement and the expressibility of the circuits is closely related with the topologies. However, the connection between these two quantities was not established hitherto. To try to understand the behavior of the average entanglement in these circuits and compare with the generation of uniformly distributed states, Fig. \ref{fig:entanglement_subtraction_ansatz1} presents a comparison between the circuits mean entanglement and the CUE analytically obtained values, normalized due to the dependency with the dimension. Excluding the case of $3$ qubits that already for $2$ layers presents values very close to zero, it is possible to observe a hierarchy between the topologies where some of them generate entanglement closer to the random values than others. This hierarchy is the same as the observed for the expressibility evolutions at higher dimensions, i.e., circuits generating average entanglement closer to the CUE will also present a steeper evolution of the expressibility at higher dimensions. This is an indicative of a connection, however it does not explain why the same hierarchy is not observed in the expressibility for smaller dimensions, whose evolutions for the connected topologies is practically the same.

\begin{figure}[htb]
    \centering
    \includegraphics[width=\columnwidth]{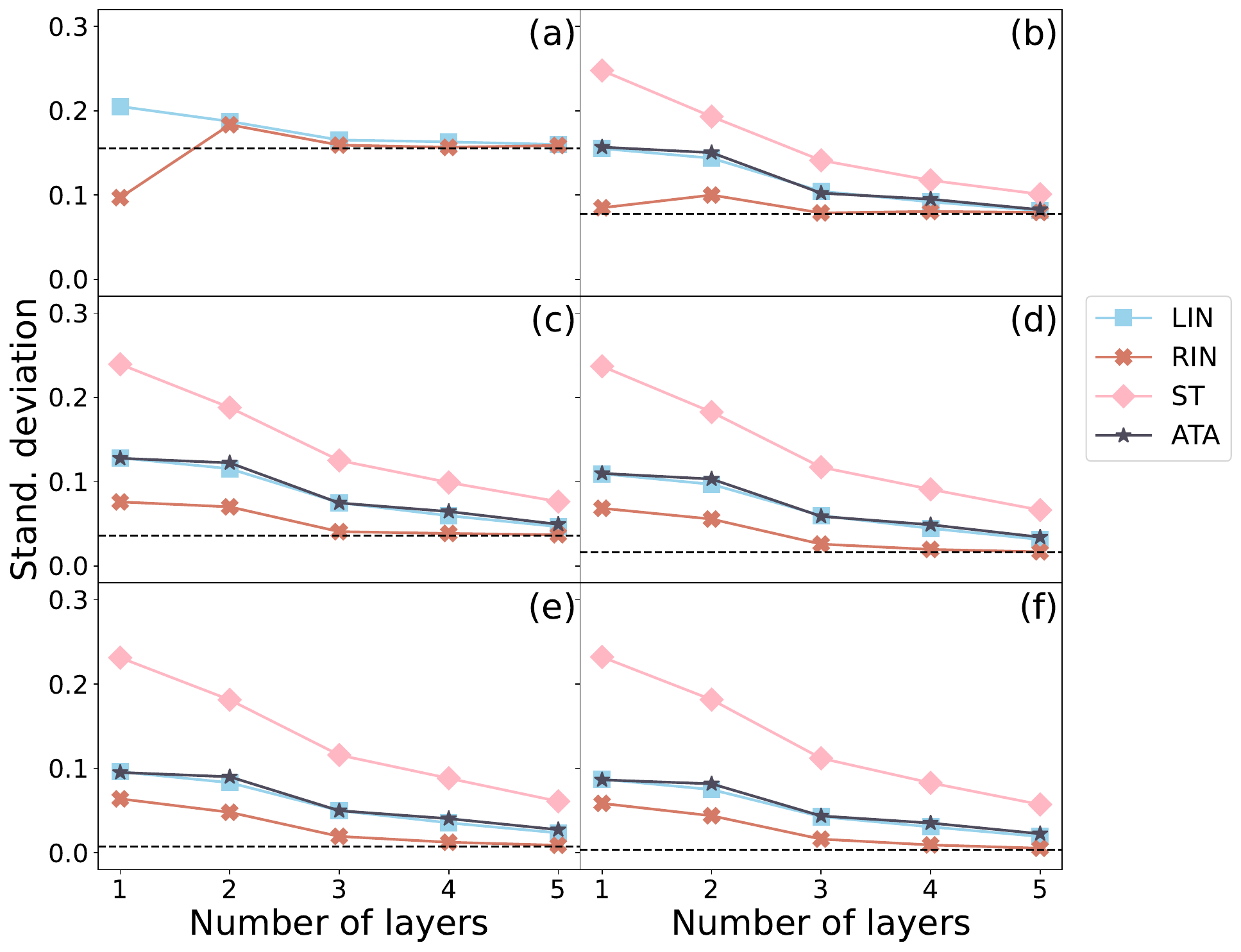}
    \caption{Ansatz 1: standard deviation of the Meyer-Wallach/Scott 1 for all connected topologies as a function of the number of layers for (a) 3 qubits, (b) 4 qubits, (c) 5 qubits, (d) 6 qubits, (e) 7 qubits and (f) 8 qubits. The traced line indicates the analytical values as obtained in Eq. \eqref{eq:analyticalaverage_CUE}. LIN: Linear, RIN: Ring, ST: Star and ATA: All-to-all.}
    \label{fig:entanglement_standdev_ansatz1}
\end{figure}

\begin{figure}[htb]
    \centering
    \includegraphics[width=\columnwidth]{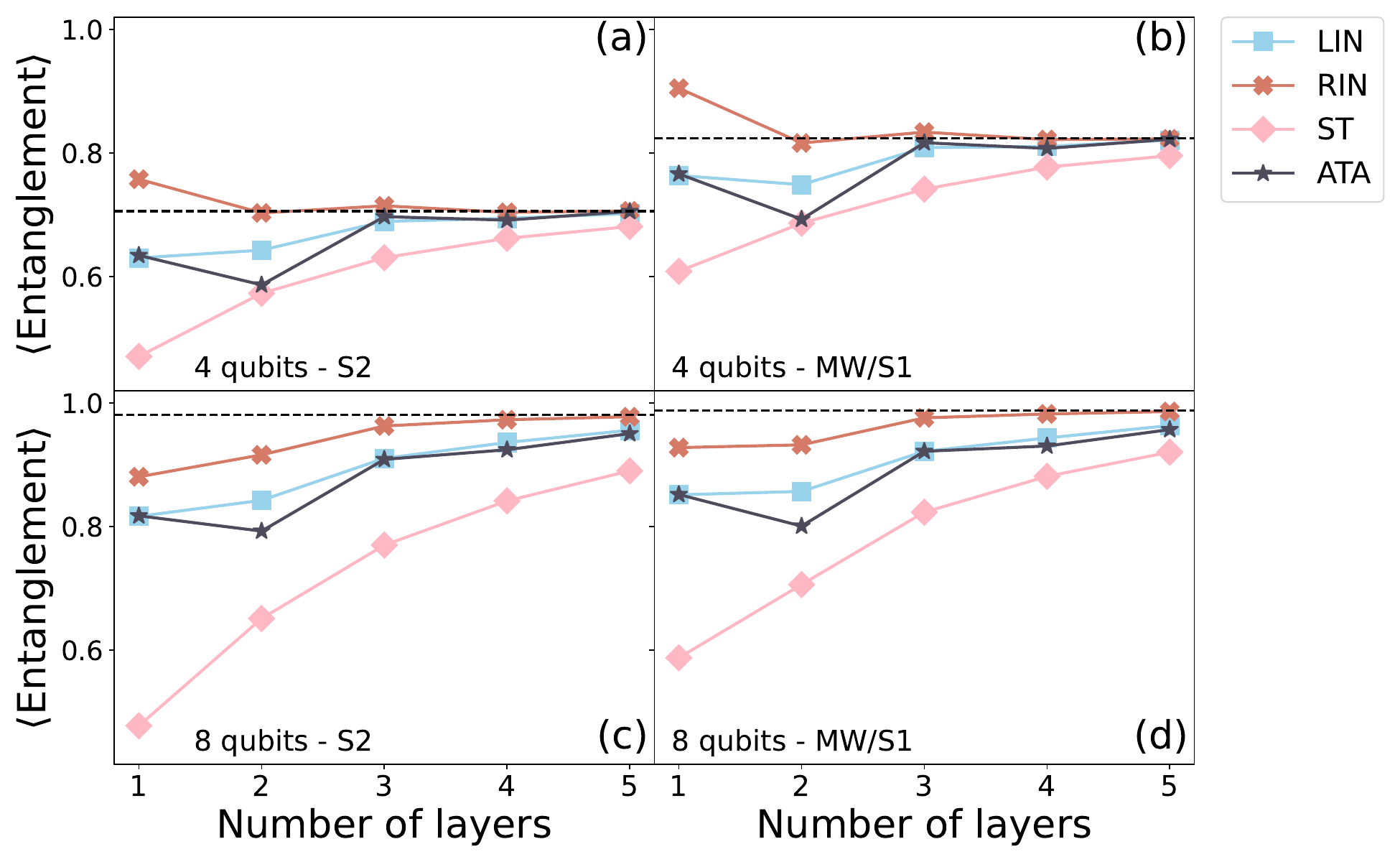}
    \caption{Comparison of the average entanglement generated by the different topologies considering the two entanglement quantifiers S1 and S2 presented in this work, for the dimensions of (a) 4 qubits - S2, (b) 4 qubits - MW/S1 and (c) 8 qubits - S2, (d) 8 qubits - MW/S1. LIN: Linear, RIN: Ring, ST: Star and ATA: All-to-all.}
    \label{fig:entanglement_comparison_ansatz1_S1_S2}
\end{figure}

To discuss this behavior in terms of another quantity, Fig. \ref{fig:entanglement_standdev_ansatz1} presents the standard deviation of the Meyer-Wallach. We can observe that, apart the $3$ qubits case, the circuits standard deviations follow the same hierarchy for the distance to the analytical values obtained from Eq.\eqref{eq:analyticalaverage_CUE} and indicated by the traced line. The Ring always generates the closest values, followed by Linear/All-to-all and the Star. However, looking at the analytical values, a characteristic of the standard deviation of slowly varying functions sampled considering uniformly random distributed states stands out. Increasing the dimension of the system, the values decay exponentially to zero, therefore for high dimensional systems the random states will present high entanglement (see Fig. \ref{fig:entanglement_comparison_ansatz1_S1_S2}) that is concentrated around the mean. This phenomenon is well known and characterized by Levy's lemma \cite{genericentanglement_concentration_measure, concentration_of_measure_barrenplateaus}, being also called concentration of measure. Due to this fact, there is a lot more freedom on the entanglement values a random state in low dimensional systems can have. However, when increasing the dimension, the random state will have very specific values for the entanglement, being these values very close to a maximally entangled state. From this, we can explain why the dimension is very influential for the evolution of the expressibility. Even if the circuit can generate entanglement values that are high and close to the analytical uniformly random values, it will only affect the evolution of the expressibility when the freedom is restricted, therefore generating the right amount and distribution of entanglement is a necessary condition for a steep evolution. This way, we can affirm that the entanglement generation for only $1$ layer, whose values have strong dependency on the topology, is influential for the evolution of the expressibility. Here, we have to take into account that the circuit analysed is Ansatz $1$, that presents many limitations in the parameterization and differences from Ansatz $2$ that will be discussed in next section.

One possible argument that could in principle explain this conclusion is that the circuits are generating average and standard deviation of entanglement closer to the analytical uniformly distributed states because they are already generating states more uniformly distributed for only $1$ layer. Then, the expressibility and statistical moments of entanglement are trivially connected. However, we can provide two counterarguments against this argument. The first one is the observation that in Fig. \ref{fig:expressibility_ans1_numberofqubits} (a), for only $1$ layer, there is no hierarchy between the circuits and this includes even the No Connections case. Therefore, there is no circuit that is explicitly and consistently closer to the Haar case in only $1$ layer of Ansatz $1$ and the entanglement generation is a fruit of the topology of connections.
From another perspective, to obtain the average value for the uniform case of the Meyer-Wallach the complete Haar measure is not needed and a $2-$design is already sufficient, as said before. This way, to quantify how close the circuit is to a $2-$design for $1$ layer can reveal if the entanglement generation is due to the structure of the connections or due to the fact that the circuit generated states are more uniformly distributed. A quantifier for how close a parameterized quantum circuit is to a $t-$design was presented in Eq. \eqref{eq:t_design_quantifier}. This quantity was numerically estimated for $20$ independent samples, considering $t=2$ and Ansatz $1$. The results are presented in Table \ref{tab:ansatz_1_2design_tests_complete}. It is possible to observe that, within the standard deviation of the calculations, all the topologies present the same behavior and are all equivalently close to a $2-$design in only $1$ layer, independent of the dimension. Therefore, we can argue that the generation of entanglement is due to the structure of the connections and to the parameterization and not because the random unitaries related with the circuits generate a $2-$design in $1$ layer.

\begin{table*}[tbp]
    \begin{center}
        \caption{Values for the quantity $||\mathcal{A}^{t=2}||^2_{2}$ defined in Eq. \eqref{eq:t_design_quantifier}, that compares the circuit with a $2-$design, obtained for the Ansatz $1$ for different topologies and number of qubits, $n$. The mean values and standard deviations were obtained over $20$ independent calculations.}\label{tab:ansatz_1_2design_tests_complete}%
        	\centering{%
        	\begin{tabular}{ccccccc}
        		\toprule
        		Topology & $n=3$ & $n=4$ & $n=5$ & $n=6$& $n=7$ & $n=8$\\
        		\midrule \midrule
        		No Connections &$(1.3\pm0.1)\cdot10^{-2}$ &$(6.5\pm0.5)\cdot10^{-3}$ &$(3.0\pm0.3)\cdot10^{-3}$ &$(1.2\pm0.1)\cdot10^{-3}$ &$(4.6\pm0.7)\cdot10^{-4}$ &$(1.6\pm0.3)\cdot10^{-4}$\\
        		\midrule 
        		Linear & $(1.3\pm0.1)\cdot10^{-2}$ &$(6.6\pm0.7)\cdot10^{-3}$ &$(3.0\pm0.3)\cdot10^{-3}$ &$(1.2\pm0.1)\cdot10^{-3}$ &$(4.5\pm0.6)\cdot10^{-4}$ &$(1.7\pm0.3)\cdot10^{-4}$\\
        		\midrule 
        		Ring & $(1.3\pm0.1)\cdot10^{-2}$ &$(6.7\pm0.6)\cdot10^{-3}$ &$(3.0\pm0.3)\cdot10^{-3}$ &$(1.2\pm0.1)\cdot10^{-3}$ &$(4.3\pm0.6)\cdot10^{-4}$ &$(1.7\pm0.3)\cdot10^{-4}$\\
        		\midrule 
        		Star & $-$ & $(6.8\pm0.5)\cdot10^{-3}$&$(3.0\pm0.3)\cdot10^{-3}$ &$(1.2\pm0.1)\cdot10^{-3}$ &$(4.4\pm0.6)\cdot10^{-4}$ &$(1.6\pm0.3)\cdot10^{-4}$\\
                \midrule
                All-to-all & $-$ &$(6.6\pm0.6)\cdot10^{-3}$&$(3.0\pm0.3)\cdot10^{-3}$ &$(1.2\pm0.2)\cdot10^{-3}$ &$(4.5\pm0.8)\cdot10^{-4}$ &$(1.6\pm0.3)\cdot10^{-4}$\\
                \bottomrule
        		\end{tabular}%
        }
    \end{center}
\end{table*}

These entanglement results are replicated considering the Scott $2$. Fig. \ref{fig:entanglement_comparison_ansatz1_S1_S2} presents a comparison for the dimensions of $4$ and $8$ qubits. It is possible to notice that the values are different, however the results are the same: The same hierarchy is observed, a very similar evolution and saturation as a function of the number of layers happen and the same conclusions can be drawn.

\subsection{\label{subsec:ansatz2}Ansatz 2}

\begin{figure}[htb]
    \centering
    \includegraphics[width=\columnwidth]{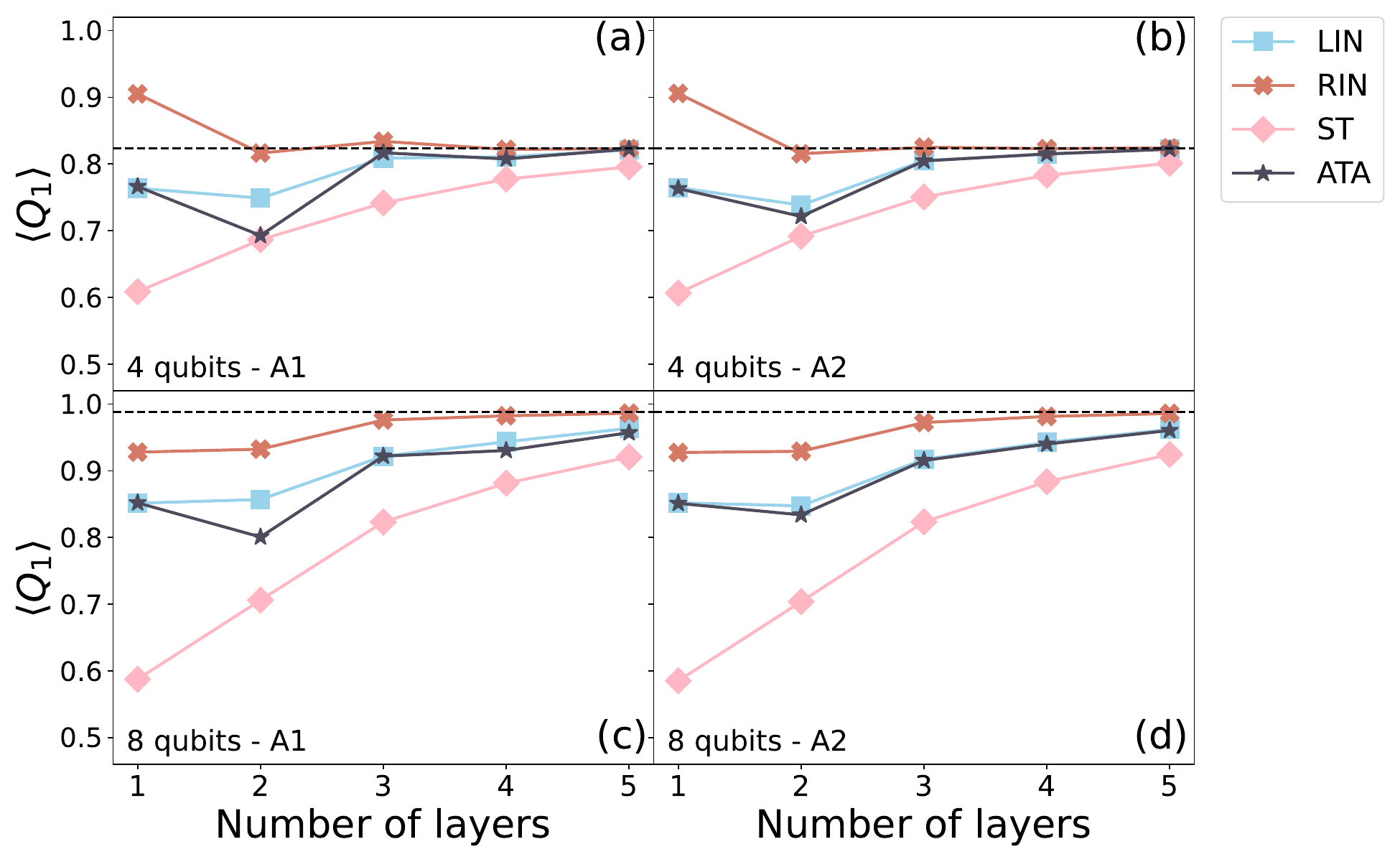}
    \caption{Comparison of the average entanglement generated by the different topologies considering the two circuit structures Ansätze $1$ (A1) and $2$ (A2) presented in this work, for the dimensions of (a) 4 qubits - A1, (b) 4 qubits - A2 and (c) 8 qubits - A1, (d) 8 qubits - A2. LIN: Linear, RIN: Ring, ST: Star and ATA: All-to-all.}
    \label{fig:entanglement_comparison_ansatz1_ansatz2}
\end{figure}

We can now move to the discussion of Ansatz $2$, starting with a comparison with Ansatz $1$. In Fig. \ref{fig:entanglement_comparison_ansatz1_ansatz2} an important result is shown: Both circuit structures present exactly the same average entanglement. This is also observed for the Scott $2$ quantifier. This way, their differences are only related with the expressibility and comes from the additional step of parameterized local operations in Ansatz $2$. In this sense, with the two structures we can isolate the role of entanglement generation and the additional freedom on the local parameters.

\begin{figure}[hbtp]
    \centering
    \includegraphics[width=\columnwidth]{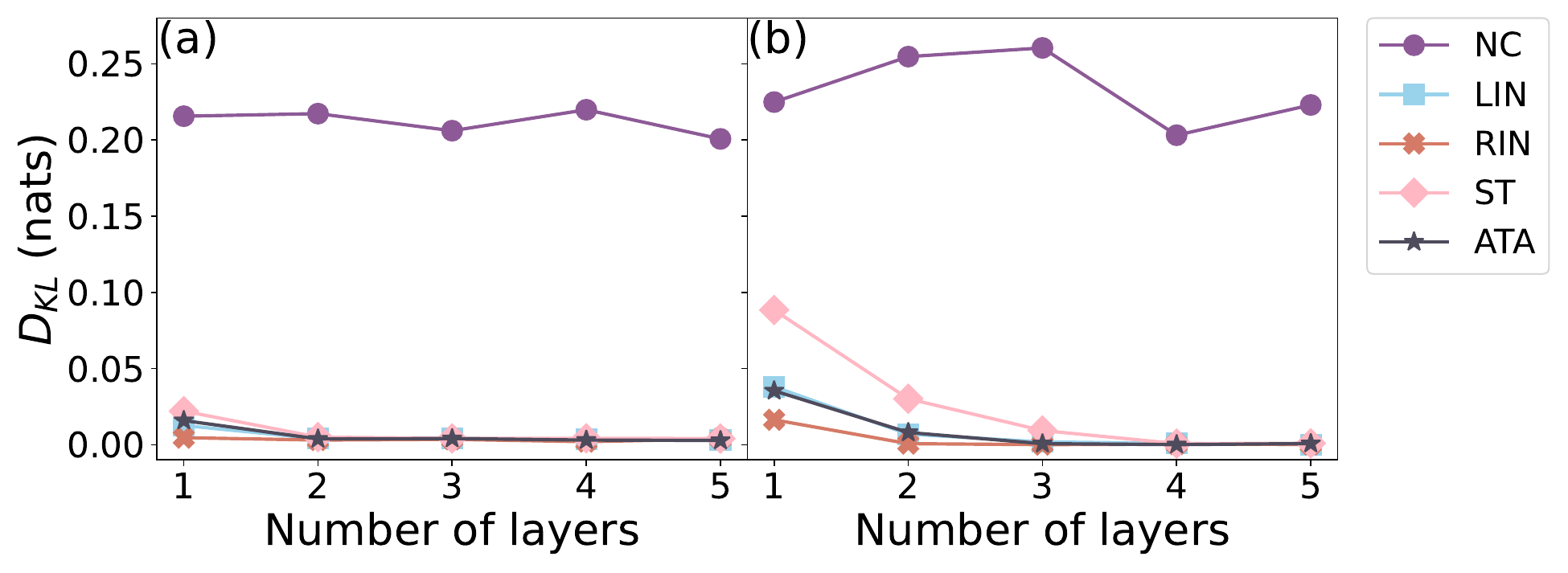}
    \caption{Ansatz 2: relative entropy as a function of the number of layers for all connectivities, considering (a) $4$ qubits and (b) $8$ qubits. NC: No Connections, LIN: Linear, RIN: Ring, ST: Star and ATA: All-to-all.}
    \label{fig:expressibility_ans2_4-8_qubits}
\end{figure}

The peculiarities of Ansatz $2$ start to appear at Fig. \ref{fig:expressibility_ans2_4-8_qubits} when comparing the different topologies together with the No Connections case for $4$ qubits (Fig. \ref{fig:expressibility_ans2_4-8_qubits} (a)) and $8$ qubits (Fig. \ref{fig:expressibility_ans2_4-8_qubits} (b)). We can see a way quicker evolution of the relative entropy for all topologies when considering Ansatz $2$ structure. The No Connections case presents the saturation values observed for Ansatz $1$ already at $1$ layer, while the connected topologies are closer to zero when compared with the No Connections. By the introduction of additional local parameters, it was possible to increase the expressibility of the circuits. Here, the connected and not connected topologies do not have the same expressibilities in only $1$, in contrast to what happens to Ansatz 1 as we have discussed in last section. This is explained by the characteristics of Ansätze $1$ and $2$. The entangled states generated with Ansatz $1$ in $1$ layer have a direct connection between entanglement and local coherences created by the RX-RY 
sequence of gates. This is due to the fact that the generation of entanglement in the CNOTs part of the circuits is only possible because of the different local coherences created by the parameterized gates. After that, entanglement and local superpositions are intrinsically connected. For Ansatz $2$, the additional step of local parameterized operations permits the manipulation of local superpositions withouth changing entanglement, therefore for every entanglement value generated after the connections part, many possible different states can be built using this different circuit structure.

\begin{figure}[htb]
    \centering
    \includegraphics[width=\columnwidth]{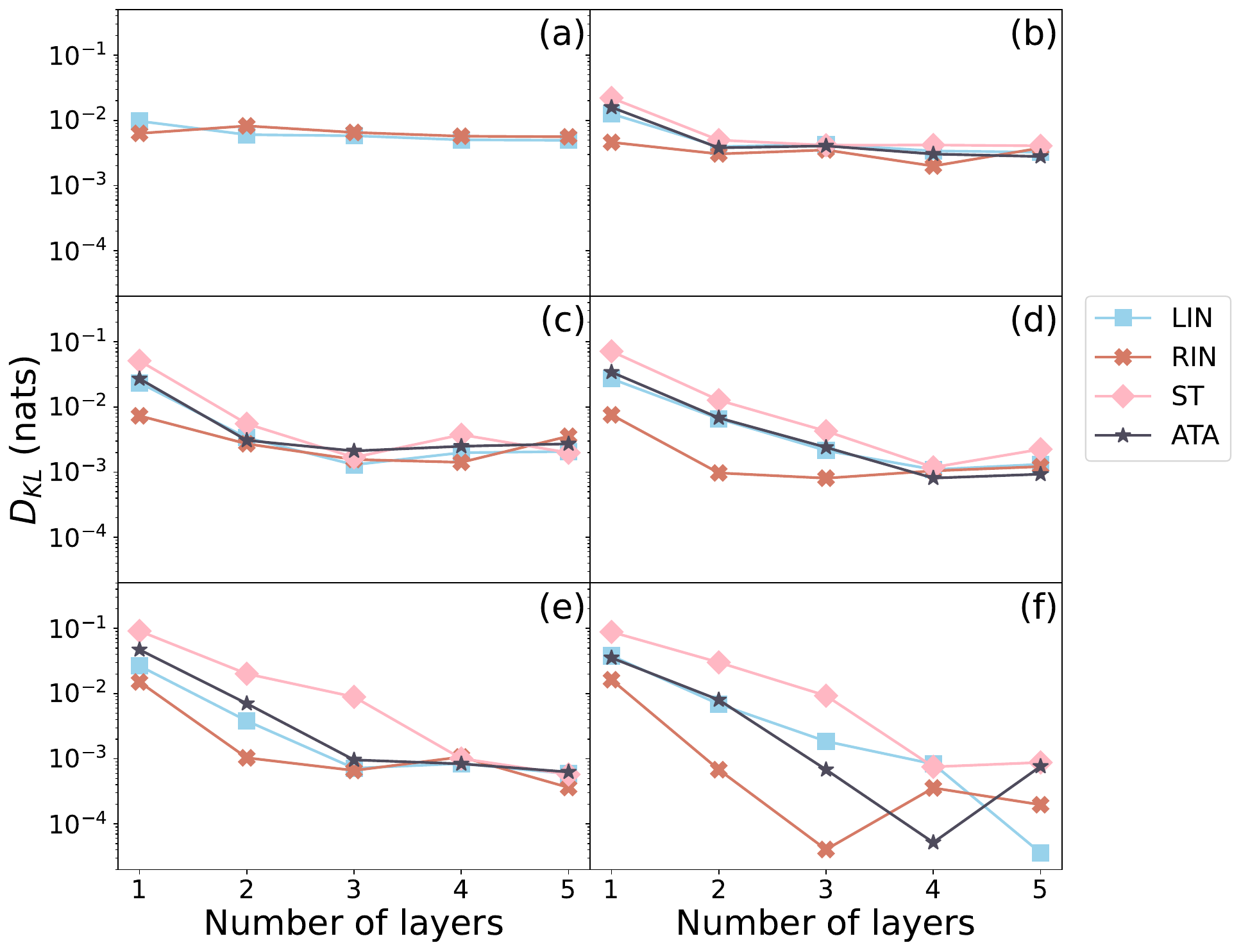}
    \caption{Ansatz 2: relative entropy as a function of the number of layers for the connected topologies, considering (a) $4$ qubits and (b) $8$ qubits. LIN: Linear, RIN: Ring, ST: Star and ATA: All-to-all.}
    \label{fig:expressibility_ans2_logscale}
\end{figure}

Now, a detailed study of the characteristics of the connected topologies is made looking at Fig. \ref{fig:expressibility_ans2_logscale}. The relative entropy evolves in a way similar to the values observed for Ansatz $1$, however requiring less layers to achieve the same results. For example, in the case of $4$ qubits (Fig. \ref{fig:expressibility_ans2_logscale} (b)), the saturation value is observed already for $2$ layers, while for Ansatz $1$ it is only observed after $4$ layers. Also, a saturation is now observed in the $6$ qubits (Fig. \ref{fig:expressibility_ans2_logscale} (d)) around $1\cdot10^{-3}$ nats, while the same case for Ansatz $1$ have a decreasing relative entropy in $5$ layers and values above $1\cdot10^{-3}$. In this sense, we again notice that Ansatz $2$ is better at achieving high expressibilities when compared to Ansatz $1$, as it has a steeper relative entropy decrease and it applies less entangling gates at the expense of additional local parameterized gates. Up to $3$ layers, the Ansatz $2$ overperforms Ansatz $1$. However, with increased number of layers, the evolutions can get erratic and there are cases where the relative entropy grows even increasing the number of layers, as Fig. \ref{fig:expressibility_ans2_logscale} (d) and (f) show, for example.

In the discussion about Ansatz $1$, it was said that the effects of generating average and standard deviation of entanglement are influential only for higher dimensions, being almost imperceptible for $4$ and $5$ qubits. Here, for Ansatz $2$, we can see the hierarchy for smaller dimensions in $1$ layer and it is well distinctive. This result may seem to contradict the previous discussion, however we have to take into account the elements present in Ansatz $2$ and not in $1$. The topologies enjoying the entanglement characteristics close to the CUE will have the additional freedom for generating entangled states with random coherences in Ansatz $2$ and we can see that this is important for the high expressibility of the circuits. Still, we can observe that this hierarchy gets more pronounced for higher dimensions and the conclusions are the same as observed for Ansatz $1$.

The results in this work find agreements with previews works analysing the expressibility and ``entangling capability'' of parameterized quantum circuits \cite{express_entang_capab}. In \cite{express_entang_capab}, there are circuits that generate average entanglement very close to the CUE mean in $1$ layer, although with relatively small expressibility when compared to circuits generating entanglement values farther from the CUE mean. There, they say that generating entanglement close to the uniformly distributed states case is a possible diagnosis for proximity to the Haar case, however it does not necessarily imply that. This is indeed true, however, they do not compare circuits with the same characteristics in terms of the gates applied, as some of them apply parameterized two qubits gates while others are only composed of not parameterized two qubits gates (CZ and CNOT, for example), which can result in very different results for expressibility and, in the context of our results, these are different analysis. When looking at the evolution as a function of the number of layers presented in their supplementary material, it is possible to notice that a circuit generating average entanglement closer to the CUE case, dubbed ``Circuit 2'' there, have a steeper expressibility increase as a function of the number of layers when compared to the others. A result also observed and systematized in our work.

\subsection{\label{subsec:limitations}{Potential limitations of the analysis}}

{In our work we decided to follow a particular reasoning to choose the particular gates performed and connectivities between qubits, however there are biases and limitations of these choices.}

{We chose circuits only with local parameterized gates and CNOTs for connections between qubits. This allowed us to isolate the influences of either connectivity or parametrization, which was important to understand how ansätze $1$ and $2$ provide different answers to the generation of entanglement and expressibility. On the contrary, parameterized two qubits gates would be an interesting choice if the objective is to maximize the expressibility with circuits of low depth, as exemplified by Ref. \cite{express_entang_capab}, whose circuits only with CNOTs as two qubits gates present a slower decay of the $D_{KL}$. On top of that, CNOT gates are unidirectional asymmetric gates, which only allow to think about topology in a non-directional graph sense. Other promising analyses, more representative and more general, would be to consider CNOT gates in both the directions of each edge of the graph or the application of symmetric Controlled-Z gates, as done in Refs. \cite{Kim_entanglementdiagnostics_VQA_randomcircuits, Kim_DarioRosa_chaosandcircuitparameters_randomcircuit}.}

{Inspired by Ref. \cite{express_entang_capab}, we considered the uniform distribution for each parameter to understand an unbiased behavior of the circuits with respect to the parameter choices: each parameter is as probable as any other within the interval. This is a reasonable choice when the lack of a cost function and a specific problem is considered, however it does not illustrate the general behavior of the circuit. Changing the distribution would affect the behavior observed for the expressibility and average entanglement, a priori. This is well pictured by the example given in the discussions, where the uniform distribution of one qubit states can be replicated considering RX and RY rotations if one considers the $2-$sphere inspired probability distribution. The same is not possible only with the two rotations and the uniform distribution. Furthermore, different types of chaotic dynamics emerge from different gates choices, as discussed in Refs. \cite{Kim_entanglementdiagnostics_VQA_randomcircuits, Kim_DarioRosa_chaosandcircuitparameters_randomcircuit}, where either the eigenvalues distributions of random matrices following the Gaussian Orthogonal Ensemble or the Gaussian Unitary Ensemble are generated by the related circuit structures.}

{When it comes to the real hardware of quantum computers usually not only one topology is available, being the circuits built with compositions of the topologies presented in this work \cite{topology_topgen_circuitgenerator, topology_superconductingqubits}. For example, in the IBM\textunderscore Ithaca quantum computer \cite{topology_topgen_circuitgenerator}, it is possible to build circuits with $12$ qubits in a Ring topology, $4$ qubits in a Star topology, many configurations of qubits in a Linear topology, and, of course, any possibility of No Connections circuits. This way, we analysed only subgraphs of the more general graph that is going to be the circuit. One work that considers this kind of possibility and discusses the consequences of a combination of the All-to-all and Ring cases is Ref. \cite{blockring_topology_2023}. From another viewpoint, the topologies will be implementable in different settings depending on the quantum computer available. For instance, ion trap quantum computers have a large freedom on the implementation of connections \cite{topology_iontraps_alltoall}, being able to perform even the most complex All-to-all topology. Nevertheless, not every hardware of this kind will be able to implement the gates we chose in this work natively. For instance, the ion trap discussed in Ref. \cite{iontrap_small_quantumcomp_atomicqubits}, which implements two qubit operations between every possible pair of qubits, can only implement RXX, RYY, and RZZ \footnote{These gates are defined as RII$(\theta):=\exp\left(-i\frac{\theta}{2}\sigma_I\otimes\sigma_I\right)$, being $\sigma_I$ the Pauli matrices with $I=X,Y,Z$.} and one qubit rotations natively. A composition of these gates would be necessary to implement the CNOT, increasing the circuit size.}

\section{\label{sec:conclusion}Conclusions}

In this work, we studied the characteristics of the generation of states in different parameterized quantum circuits whose structures were chosen based in contemporary quantum computer architectures. Five different connectivities between qubits were analysed, together with two different parameterizations, generating ten different circuits consisting of parameterized one qubit X and Y rotations and CNOT gates. To characterize how uniformly distributed are the random states and what are the average and standard deviation of entanglement values, we applied the expressibility quantifier together with the averages of Scott entanglement quantifier of order $1$ and $2$. The results showed that circuits with a Ring topology will have the steepest increase of the expressibility as a function of the number of layers and highest entanglement values, followed by Linear/All-to-all practically together and finally the Star circuit. The No Connections circuits do not generate entanglement and will have saturated expressibility values.

These results are influenced by the dimension of the system, however the same trend is observed. An important correlation is noticed: Circuits generating entanglement averages and standard deviation deviation closer to the circular unitary ensemble, that is uniformly distributed over the unitary manifold, have a steeper increase of the expressibility. This is justified in terms of the freedom of the random states that can be produced depending on the dimension of the system, and generating states with closer average and standard deviation of entanglement favors the increase of expressibility.

The proper entanglement generation and its advantage is explored with the two possible parameterizations. In Ansatz $1$, only one step of local parameterized operations is applied before the connections between qubits, while in Ansatz $2$ the connections between qubits is interleaved with local parameterized operations. This leads to a quicker increase of the expressibility as a function of the number of layers, maintaining the entanglement constant. Therefore, by applying more local unitary operations, that are less complex and less costly when compared to two qubit operations, it was possible to obtain a more uniform distribution of states. The results were replicated with both entanglement measures.

The Meyer-Wallach/Scott quantifiers have been applied to characterize random circuits and chaotic systems for decades now \cite{Viola_parameters_of_pseudorandomcircuits, scott_baker_map, Scott_entanglement_measure}. For example, in Ref. \cite{Viola_parameters_of_pseudorandomcircuits}, the Meyer-Wallach is applied as a possible diagnosis of the convergence of circuits with different coupling topologies and local gates to a pseudorandom circuit. Here, we apply the same entanglement measure together with Scott$-2$ and the entanglement generation is not a diagnosis of convergence as in these previous works, but is applied as a way to understand how the different entanglement generation of the circuits influences the convergence to a uniformly distributed case, quantified by expressibility.

{Since the introduction of the expressibility and average entanglement quantifiers as defined in Ref. \cite{express_entang_capab}, no other work investigated the connections between the entanglement standard deviation of the circuits at only $1$ layer and the expressibility increase as a function of circuit depth. The average entanglement growth and standard deviation decay as a function of the number of layers together with the randomness increase is a known characteristic of random circuits, and has been appreciated in many important works \cite{scott_baker_map, Scott_entanglement_measure, Viola_parameters_of_pseudorandomcircuits, genericentanglement_concentration_measure,entanglement_production_PQC_mangini, concentration_of_measure_barrenplateaus, entanglement_induced_barrenplateaus}. Still, the efforts to understand randomness and entanglement generation in parameterized quantum circuits are even more recent \cite{randomness_PQCs_ensembles, noisy_PQCs_random, entanglement_production_PQC_mangini}. Leading results have investigated how circuits with this kind of structure will be able to generate $t-$designs \cite{randomness_PQCs_ensembles, concentration_of_measure_barrenplateaus}, will have an increase of entanglement that presents a universal increase independent on the number of qubits \cite{entanglement_production_PQC_mangini}, a result also observed here, and finally how this randomness generation is influenced by the introduction of noise \cite{noisy_PQCs_random}. Recently, how the randomness in PQCs can be compared to other kinds of random circuits \cite{optimal_complexity_PQCs} has been explored, including analysis applying the majorization criterion for complexity quantification. Parameterized quantum circuits are one of the essential ingredients of Variational Quantum Algorithms, and the efforts to extend and understand their possibilities are vast, including results for quantum batteries and thermodynamics \cite{tai_VQA_ergotropy, variational_quantum_thermalizer}, reproduction of search algorithms in the NISQ era \cite{search_algorithm_VQA}, and Quantum Machine Learning \cite{groupinvariant_QML, variational_quantum_classifier}. Ref. \cite{Cerezo2021VqaReview} provides an extensive review of the many possible applications of PQCs in VQAs.}

This work provides results that can be applied for choosing quantum architectures depending on the characteristics expected from the circuits and places perspectives on the possibility of application of parameterized quantum circuits for the generation of random states. For example, if the circuit requires high entanglement values and/or distribution of states close to the uniform, an interesting candidate would be circuits with Ring topology. Conversely, if the intention is to have circuits with less entanglement and/or less uniformly distributed states, the Star would be the best choice between the options we studied. The circuits Linear/All-to-all are the middle point between the possibilities. Still, the expressibility quantifier is not the unique choice and other circuit complexity quantifiers would be interesting to study these characteristics and provide deeper insights in the behavior of parameterized quantum circuits and their generation of random states.

{The results here place a connection between entanglement generation and complexity of circuits as quantified by the expressibility. A natural extension of this discussion is how these can be connected to notions of information scrambling, where local information in the input system cannot be accessed locally in the output \cite{chaos_in_quantum_channels}. Whether or not the same behavior is observed, i.e., circuits that are more scrambling after one layer present a steeper increase of expressibility, the relations between entanglement and information scrambling in this setting would benefit from the insights provided. From another viewpoint, the findings could be analysed in comparison with the performance of specific cases of VQA applications, for instance the Variational Quantum Eigensolver search for ground states, to understand which are the consequences to trainability. A result already observed is that circuits whose topology is completely different from the connections observed in the Hamiltonian will have a worse performance when compared to circuits with topology following the Hamiltonian structure \cite{ivanmedina2023vqeinspired}. However, how this connects to the average entanglement and expressibility is still not clear. Examples of this approach are presented in Refs. \cite{Kim_entanglementdiagnostics_VQA_randomcircuits,Kim_DarioRosa_chaosandcircuitparameters_randomcircuit}. As discussed in Sec. \ref{subsec:limitations}, the circuits here do not include all potential circuits appearing in quantum computers. In this sense, another promising extension of this work would be to understand how the use of different topologies in the same quantum circuit will affect the behavior of the expressibility and entanglement generation. Finally, bounds on the entanglement of quantum states were already applied to determine the possible classical simulability of quantum dynamics \cite{vidal_simulation_entangled_states}, and the complexity of quantum circuits was shown to be bounded below by their entangling power \cite{eisert2021entangling}. In this sense, the observed connection between expressibility and average entanglement at only one layer can be explored to understand circuit complexity growth in a more fundamental sense, and is an interesting research project for future works.}

\begin{acknowledgments}
This study was financed in part by the Coordenação de Aperfeiçoamento de Pessoal de Nível Superior – Brazil (CAPES) – Finance Code 001. G.I.C. acknowledges the financial support of the Research Council of Finland through the Finnish Quantum Flagship project (358878, UH). I.M. acknowledges financial support from São Paulo Research Foundation - FAPESP (Grant No. 2022/08786-2 and No. 2023/14488-7). P.C.A. acknowledges financial support from Conselho Nacional de Desenvolvimento Científico e Tecnológico - Brazil (CNPq - Grant No. 160851/2021-1). D.O.S.P acknowledges the support by the Brazilian funding agencies CNPq (Grants No. 304891/2022-3 and No. 402074/2023-8), FAPESP (Grant No. 2017/03727-0) and the Brazilian National Institute of Science and Technology of Quantum Information (INCT/IQ). 
\end{acknowledgments}


%

\end{document}